\documentclass[a4paper,11pt]{article}
\usepackage{jheppub} 
\usepackage{cleveref}

\usepackage{graphicx}
\usepackage{slashed}
\graphicspath{{figures/}}
\usepackage{subcaption}
\usepackage{bbold}
\usepackage{multirow}
\usepackage{makecell}
\usepackage{xcolor}

\allowdisplaybreaks[4]

\title{Neutrinoless double beta decays of hyperons in covariant chiral perturbation theory}
\author[a,b]{Zi-Ying Zhao}
\emailAdd{zhaoziying@hnu.edu.cn}
\author[c,a]{Ze-Rui Liang}
\emailAdd{liangzr@hebtu.edu.cn}
\author[d,e]{Feng-Kun Guo}
\emailAdd{fkguo@itp.ac.cn}
\author[f]{Li-Ping He}
\emailAdd{heliping@csu.edu.cn}
\author[a,b]{De-Liang~Yao}
\emailAdd{yaodeliang@hnu.edu.cn}

\affiliation[a]{School of Physics and Electronics, Hunan University, Changsha 410082, China}
\affiliation[b]{Hunan Provincial Key Laboratory of High-Energy Scale Physics and Applications, Hunan University, Changsha 410082, China}
\affiliation[c]{College of Physics and Hebei Key Laboratory of Photophysics Research and Application, Hebei Normal University, Shijiazhuang, Hebei 050024, China}
\affiliation[d]{Institute of Theoretical Physics, Chinese Academy of Sciences, Beijing 100190, China}
\affiliation[e]{School of Physical Sciences, University of Chinese Academy of Sciences, Beijing 100049, China}
\affiliation[f]{School of Physics, Central South University, Changsha 410083, China}

\abstract{Neutrinoless double beta ($0\nu\beta\beta$) decays of spin-1/2 hyperons are investigated in a covariant baryon chiral perturbation theory framework, extended by a $\Delta L=2$ operator proportional to the Majorana neutrino mass, where $L$ denotes the lepton number. Within the light Majorana neutrino exchange mechanism, the decay amplitudes are found to emerge at the one-loop level, representing the long-range contribution. The extended-on-mass-shell scheme is employed to renormalize the one-loop amplitudes and restore consistent chiral power counting. Consequently, the differential decay rates for all accessible hyperon $0\nu\beta\beta$ channels are predicted and the corresponding branching ratios are more than 20 orders of magnitude smaller than the current experimental upper bounds. Interestingly, it is found that the leading contribution to hyperon $0\nu\beta\beta$ decay is actually from short-range counterterm operators, as required by the renormalization argument. Neutrinoless transition form factors are proposed to determine this leading contribution through future lattice QCD simulations.
}
\keywords{Neutrinoless double beta decay, hyperon decays, lepton number violation,
baryon chiral perturbation theory, Majorana neutrinos}

\begin{document}
\maketitle
\flushbottom

\section{Introduction}
\label{sec:intro}

The search for neutrinoless double beta ($0\nu\beta\beta$) decay is one of the most promising approaches to probe lepton number violation (LNV) and to test the Majorana nature of neutrinos~\cite{Schechter:1981bd}; see, e.g., refs.~\cite{Rodejohann:2011mu,DellOro:2016tmg,Dolinski:2019nrj} for comprehensive reviews. A positive observation would be definitive evidence for the Majorana nature of neutrinos, with profound implications for constraining the neutrino mass hierarchy and absolute scale. Nuclear $0\nu\beta\beta$ decay is currently the primary method for searching for LNV, extensively measured in various experiments~\cite{IGEX:2002bce,Danevich:2003ef,Umehara:2008ru,EXO-200:2019rkq,GERDA:2020xhi,KamLAND-Zen:2022tow,XENON:2022evz,CUPID:2022puj,Majorana:2022udl,CUPID-Mo:2022cel,NEXT:2023daz,PandaX:2023ggs,KamLAND-Zen:2024eml,Garfagnini:2024rvs,CUORE:2024ikf}. However, the calculation of its half-life depends sensitively on the nuclear matrix element, which varies significantly among different nuclear models, leading to discrepancies in predicted half-lives by up to an order of magnitude~\cite{Engel:2016xgb,DellOro:2016tmg,Dolinski:2019nrj}. 

As a complementary avenue, LNV processes in hyperon decays offer an alternative way to search for LNV at the hadronic level. The hyperon $0\nu\beta\beta$ decays involve at least a change in strangeness and correspond to flavor-changing weak processes, such as $s \to u$ and $d \to u$ transitions (see e.g., refs.~\cite{Littenberg:1991rd} for early discussions), which possess a typical strength of second-order weak decays. Although hyperon decay experiments cannot reach the extremely small branching fractions that are characteristic of nuclear decays, they provide access to decay modes with muons in the final state, which are forbidden in nuclear environments due to limited phase space. Furthermore, it is theoretically much easier to determine the hyperon decay amplitude than the nuclear matrix element. Therefore, hyperon $0\nu\beta\beta$ decays constitute an ideal and complementary probe in the broader search for LNV signal. 

Several experimental searches have been conducted to constrain LNV hyperon decays~\cite{Li:2016tlt,Goudzovski:2022vbt}. The first upper bound on the $\Delta L=2$ hyperon decay, $\mathcal{B}(\Xi^- \to p\mu^-\mu^-)< 3.7\times10^{-4}$ at $90\%$ confidence level (C.L.), was established in ref.~\cite{Littenberg:1991rd} through a retroactive analysis of the $\Xi^-$ rare decay data collected at BNL~\cite{Yeh:1974wv}. This limit was updated by the HyperCP Collaboration to $\mathcal{B}(\Xi^- \to p\mu^-\mu^-)< 4.0\times10^{-8}$ at $90\%$ C.L.~\cite{HyperCP:2005sby}. 
The BESIII Collaboration reported that $\mathcal{B}(\Sigma^- \to pe^-e^-)< 6.7\times 10^{-5}$ and $\mathcal{B}(\Sigma^- \to \Sigma^+X)< 1.2\times 10^{-4}$ at $90\%$ C.L.~\cite{BESIII:2020iwk}. More recently, BESIII has further improved the experimental sensitivity by setting a new upper bound on $\mathcal{B}(\Xi^- \to \Sigma^+ e^-e^-)< 2\times 10^{-5}$ at 90\% C.L.~\cite{BESIII:2025ylz}.  Bounds on hyperon $0\nu\beta\beta$ decays are expected to be further improved through measurements carried out at future facilities, such as the Super Tau-Charm Facility (STCF)~\cite{Achasov:2023gey}, with a two-orders-of-magnitude higher luminosity. We note that far richer data on LNV decays of charged $B$, $D$, and $K$ mesons have accumulated~\cite{Appel:2000tc,BELLE:2011bej,LHCb:2013hxr,BaBar:2013swg,LHCb:2014osd,NA62:2019eax}, compared to those for hyperon LNV decays. The existing experimental upper limits on the corresponding branching ratios of hyperon decay channels provide direct constraints on the underlying LNV dynamics and can be translated into bounds on the effective Majorana neutrino mass within a given theoretical framework. Conversely, for decay modes for which no experimental limits are currently available, the neutrino mass bounds inferred from nuclear $0\nu\beta\beta$ experiments can be employed to estimate the hyperon $0\nu\beta\beta$ branching ratios. 

Theoretical investigations of LNV hyperon decays are relatively limited. Early studies, such as ref.~\cite{Barbero:2007zm}, estimated branching ratios within the framework of virtual Majorana neutrino exchange and obtained $\mathcal{B}(\Sigma^- \to p e^-e^-)\approx 1.49\times10^{-35}$, assuming an effective Majorana neutrino mass of $10~\rm eV$. By extending the theoretical framework to include local six-quark $\Delta L=2$ operators alongside light-neutrino exchange, ref.~\cite{Barbero:2013fc} evaluated the relevant hadronic matrix elements within the bag model, yielding a prediction of $\mathcal{B}(\Sigma^- \to p e^-e^-)\leq 10^{-23}$. To obtain refined estimates, the calculation of hyperon $0\nu\beta\beta$ decays~\cite{Barbero:2007zm,Barbero:2013fc} was recently revisited in ref.~\cite{Hernandez-Tome:2021byt}, where the loop divergence is tamed by form factors and the short-range contribution is estimated by incorporating heavy new-physics degrees of freedom. Despite the widespread applications of modern effective field theory (EFT) to both nuclear~\cite{Cirigliano:2017djv,Cirigliano:2017tvr,Cirigliano:2018hja} and meson~\cite{Liao:2019gex,Liao:2020roy,Zhou:2021lnl,Chen:2025svf} $0\nu\beta\beta$ decays, the hyperon sector remains unexplored. 

In this work, we fill this gap by performing a systematic calculation of $0\nu\beta\beta$ decays of spin-$1/2$ hyperons within SU(3) baryon chiral perturbation theory (BChPT)~\cite{Weinberg:1978kz,Gasser:1984gg,Bernard:1995dp}, the EFT of QCD at low energies. To this end, the standard chiral effective Lagrangian~\cite{Gasser:1983yg,Georgi:1984zwz,Krause:1990xc,Oller:2006yh} is enlarged by including a LNV operator, which is mapped from the dimension-$5$ Weinberg operator~\cite{Weinberg:1979sa}, responsible for the generation of Majorana neutrino mass. In general, the dynamical $0\nu\beta\beta$ mechanism can be classified into two categories~\cite{Cirigliano:2017djv}: long-range (involving neutrino exchange) and short-range (without neutrino exchange).
Here we are only concerned with the part in the long-range contribution from light neutrino exchange, which is typically proportional to the Majorana neutrino mass. This part is traditionally called the mass mechanism and is graphically exhibited in figure~\ref{fig.mass.mech}. The mass mechanism is fully captured by the dimension-$5$ Weinberg operator, preventing us from introducing any other higher-dimensional operators from standard model EFT (SMEFT)~\cite{Babu:2001ex,deGouvea:2007qla,Lehman:2014jma,Liao:2020jmn}. That is, the mass mechanism actually corresponds to the leading order (LO) long-range contribution, since the higher-dimensional LNV operators are suppressed by the inverse of hard scales in EFT. In addition, in the mass mechanism, the weak interaction between hadrons and leptons is mediated by the SM charged current, whereas the LNV vertex is confined to the lepton sector. 

In BChPT, the hyperon $0\nu\beta\beta$ decay driven by the mass mechanism begins to contribute at one-loop level, i.e. $\mathcal{O}(p^3)$, with $p$ collectively denoting small quantities in chiral expansion. The aforementioned feature of the mass mechanism guarantees that the one-loop decay amplitude factorizes into a product of hadronic and leptonic tensors, modulo an overall factor, as shown in eq.~\eqref{eq.amplitude.decomposition}. The ultraviolet (UV) divergence of the one-loop amplitude is subtracted by employing the dimensional regularization (DR) with the $\overline{\rm MS}-1$ subtraction scheme. Besides, the notable power counting breaking (PCB) issue~\cite{Gasser:1987rb} is remedied by further imposing the extended-on-mass-shell (EOMS) scheme~\cite{Fuchs:2003qc}. 
The EOMS scheme has been successfully applied to various quantities and processes, such as nucleon mass~\cite{Fuchs:2003qc,Liang:2025cjd,Chen:2024twu}, nucleon sigma terms~\cite{Alarcon:2012nr,Liang:2025adz}, nucleon form factors~\cite{Schindler:2006it,Yao:2017fym,Bernard:2025gto}, nucleon scatterings~\cite{Alarcon:2012kn,Yao:2016vbz,Siemens:2017opr,Yao:2018pzc,Lu:2021gsb,Lu:2022hwm,Chen:2024kbh}, baryon magnetic moments~\cite{Geng:2008mf} and heavy-light interactions~\cite{Yao:2015qia,Liang:2023scp}. 
Based on the one-loop renormalized amplitude, we compute the long-range one-loop contributions to the differential decay rates and branching ratios for six hyperon $0\nu\beta\beta$ decay processes: $(\Sigma^-,\Xi^-)\to(p,\Sigma^+)e^-e^-$ and $(\Sigma^-,\Xi^-)\to p \mu^-\mu^-$, which are allowed by kinematics. 
For the four electronic modes, the pertinent branching ratios from the light-neutrino exchange turn out to be of the order $10^{-31}$ or less. In evaluating the branching ratios, we use a
benchmark value \(m_{ee}=100~{\rm meV}\) for the effective Majorana mass. This benchmark is
chosen as a representative value consistent with current nuclear $0\nu\beta\beta$  constraints in the standard light-neutrino mass mechanism; the translation from half-life limits to $m_{ee}$ is
nuclear-matrix-element dependent (see, e.g., ref.~\cite{KamLAND-Zen:2024eml,EXO-200:2019rkq,CUORE:2024ikf,Majorana:2022udl,GERDA:2020xhi}). 
For the two muonic modes, the relevant branching ratios are of the order $10^{-28}$, estimated with a benchmark effective Majorana mass $m_{\mu\mu}= 10\, \text{eV}$. 
We also explore the behaviour of the branching ratios by varying the effective Majorana mass. 
A rough estimate, obtained by naively mapping the experimental upper limits by HyperCP~\cite{HyperCP:2005sby} and BESIII~\cite{BESIII:2020iwk,BESIII:2025ylz} onto the light-neutrino-exchange parameterization, yields the indicative scales $m_{\mu\mu}\sim 100$~GeV and $m_{ee}\sim 1$~TeV, respectively. 

It should be emphasized that the UV and PCB renormalization requires introducing extra local LNV operators as counterterms, accompanied by unknown low-energy constants (LECs). The counterterm contribution accounts for contact interactions between baryons and charge leptons, and therefore corresponds to the short-range mechanism. We find that the counterterm contribution to the decay amplitude contains an $\mathcal{O}(p^2)$ piece, lower than the $\mathcal{O}(p^3)$ one-loop contribution. 
Thus, the LO contribution to hyperon $0\nu\beta\beta$ decay is from short-range operators, as demanded by the renormalization argument. Similar finding was also observed for nuclear $0\nu\beta\beta$ decay in ref.~\cite{Cirigliano:2018hja}. Unfortunately, due to the unknown LECs, we do not include the counterterm contribution in our numerical prediction discussed in the previous paragraph. A promising approach to determine the LECs appearing in the counterterms is matching to lattice QCD. Therefore, we propose to define neutrinoless transition form factors (TFFs) for the kinematic configuration where the two final leptons carry identical momenta. These neutrinoless TFFs are suitable for lattice QCD simulations in the future. We derive the chiral expressions of the TFFs and explore their properties by varying the momentum transfer squared and the pion mass. 

The paper is organized as follows. We introduce the kinematics and the effective Lagrangians relevant for our analysis in section~\ref{sec:kinematics} and section~\ref{sec:Effective_Lagrangians}, respectively.
The generic structure of the decay amplitude is discussed in section~\ref{sec:Amp}. Section~\ref{sec:OneLoop_Amp} is devoted to the calculation of the one-loop amplitudes, including the contributions from the crossed channels. The renormalization procedure is described in section~\ref{sec:Renormalization}, where the counterterm Lagrangian is constructed. In section~\ref{sec:Brach_ratio}, we present our numerical results of the differential decay rates and branching ratios, providing predictions for the various decay modes under the current constraints on the effective Majorana neutrino mass. The TFFs are defined and numerically analyzed in section~\ref{sec:TFF}. Finally, section~\ref{sec:summary} contains our summary and outlook.
Technical details are relegated to the appendices, including reduction of the leptonic part (appendix~\ref{app:lept}), Lorentz operators suitable for chiral expansion (appendix~\ref{app:operators}), explicit one-loop expressions of the hadronic tensors (appendix~\ref{app:hadronic_Amp}), relations between the $t$- and $u$-channel structure functions (appendix~\ref{app:cross}), and explicit one-loop expressions for the neutrinoless TFFs (appendix~\ref{app:TFF_expressions}).

\section{Decay amplitudes in chiral perturbation theory}
\label{sec:thero}
\subsection{Kinematics}
\label{sec:kinematics}

\begin{figure}[tb]
\centering
\includegraphics[width=0.42\linewidth]{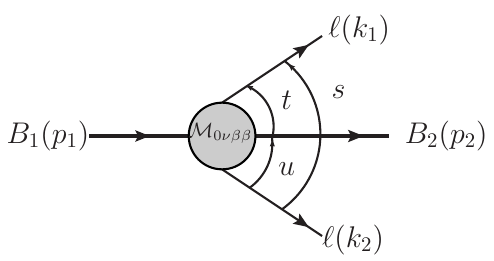} \caption{Kinematics of hyperon $0\nu\beta\beta$ decays.}
\label{fig.kinematics}
\end{figure}

For spin-$1/2$ hyperons, there are four physical processes of $0\nu\beta\beta$ decay that are allowed by kinematics, which can be classified by the change of strangeness $\Delta S$ as follows,
\begin{align}
\Delta S =0: ~& \Sigma^-[dds]\to \Sigma^+[uus]+\ell^-+\ell^-\ ,\notag \\
\Delta S =1: ~& \Sigma^-[dds]\to p[duu]+\ell^-+\ell^-\ , \quad\Xi^-[dss]\to \Sigma^+[uus]+\ell^-+\ell^-\ ,\notag \\
\Delta S =2: ~& \Xi^-[dss]\to p[duu]+\ell^-+\ell^-\ .\label{eq.physical.processes}
\end{align}
For clarity, the kinematics of the $0\nu\beta\beta$ decays are graphically shown in figure~\ref{fig.kinematics}. The decay processes can be described generically by 
\begin{align}
    B_1^-(p_1) \rightarrow B_2^+(p_2) + \ell^-(k_1) + \ell^-(k_2)\ ,
\end{align}
where $p_{1,2}$ and $k_{1,2}$ in the parentheses denote the momenta of the baryons and leptons, respectively. Here, one has $B_1^-\in\{\Sigma^-,\Xi^-\}$ and $B_2^+\in\{\Sigma^+,p\}$. The outgoing dilepton can be either electrons or muons, provided that the final-state phase space is sufficient. The Lorentz-invariant Mandelstam variables are defined by
\begin{equation}
s=(k_1+k_2)^2=(p_1-p_2)^2, \quad t=(p_2+k_1)^2, \quad u=(p_2+k_2)^2\ ,
\end{equation}
fulfilling the constraint $s+t+u=m_{1}^2+m_{2}^2+2m_{\ell}^2\equiv \Sigma_m$, with $m_1$, $m_2$ and $m_\ell$ being the masses of the initial hyperon, the final baryon and the leptons, respectively.

The partial decay rate in the rest frame of the initial hyperon $B_1$ is expressed in terms of the invariant matrix element $\mathcal{M}_{0\nu\beta\beta}$ by~\cite{ParticleDataGroup:2024}\footnote{We follow the convention $S=1-{\rm i}(2\pi)^4\delta^{(4)}(p_1-p_2-k_1-k_2)\mathcal{M}$ to define the decay amplitude $\mathcal{M}_{0\nu\beta\beta}$, where $S$ refers to the $S$-matrix element. }
\begin{align}
{\rm d}\,\Gamma =\frac{1}{(2\pi)^3}\frac{1}{32m_{1}^3}\overline{|\mathcal{M}_{0\nu\beta\beta}|^2}\,{\rm d}s\,{\rm d}u\ ,
\label{eq.partial.decay.rate}
\end{align}
where the overline stands for averaging over the spin states. The kinematically allowed range of the Mandelstam variable $s$ is \(4 m_{\ell}^2 \leq s \leq (m_{1} - m_{2})^2\). For a fixed $s$, \(u\) can take values in the range
\begin{align}
u\in[u_-,u_+]\ ,\quad
u_\pm=(\omega + E_2)^2 - \bigg[\sqrt{\omega^2 - m_{\ell}^2} \mp \sqrt{E_2^2 - m_{2}^2}\bigg]^2\ , \label{eq.u.limits}
\end{align}
where $\omega=\sqrt{s}/{2}$ and $E_2=(m_1^2-m_2^2-s)/(2\sqrt{s})$ are the energies of one of the leptons and the final baryon in the center-of-mass frame of the dilepton system.

\subsection{Effective Lagrangians}
\label{sec:Effective_Lagrangians}

The $0\nu\beta\beta$ process arises from the LNV operators of odd dimensions in the framework of SMEFT~\cite{Weinberg:1979sa,Babu:2001ex,deGouvea:2007qla,Lehman:2014jma}. Provided that the LNV scale $\Lambda_{\rm LNV}$ is much higher than the electroweak scale $\Lambda_{\rm EW}$, the only low-energy manifestation of LNV is a Majorana mass for light neutrinos, as detailed in ref.~\cite{Cirigliano:2017djv}. It is well-known that the dimension-5 Weinberg operator in the SMEFT gives rise to the Majorana mass term for the left-handed neutrinos~\cite{Weinberg:1979sa}. The $0\nu\beta\beta$ decay is then induced via light Majorana-neutrino exchange between two SM charged current vertices. This is the so-called mass mechanism of $0\nu\beta\beta$ decays as mentioned in the Introduction. 

Specifically, at the electroweak scale, the SM Lagrangian is augmented to
\begin{align}
    \mathcal{L}_{\rm eff} =\mathcal{L}_{\rm SM} + \bigg[\frac{y_{\ell\ell^{\prime}}}{\Lambda_{\rm LNV}}  \bar{L}_{\ell L} \widetilde{H}  \widetilde{H}^T L_{\ell^{\prime} L}^c  + \text{h.c.}\bigg]\ ,\label{eq.Lag.eff.SMEFT}
\end{align}
where $y_{\ell\ell^\prime}$ ($\ell,\ell^{\prime}\in \{e,\mu,\tau\}$) are dimensionless coefficients, $L_{\alpha L}=(\nu_{\ell L},\ell_{L})^T$ is the left-handed SU(2) lepton doublet, and its charge-conjugate counterpart is given by $L^c_{\ell^{\prime} L}= C \bar{L}_{\ell\prime L}^T$. Summation over repeated indices is implied. Here $C={\rm i}\gamma^2\gamma^0$ is the charge conjugation matrix with property $C=-C^{T}=-C^{\dagger}=-C^{-1}$. Furthermore, $H$ stands for the Higgs doublet and $\widetilde{H}={\rm i}\tau_2 H^\ast$ with $\tau_2$ the Pauli matrix. After spontaneous symmetry breaking, the Higgs acquires a vacuum expectation value $\langle H \rangle = (0, v/\sqrt{2})^T$, and the operator in the square brackets of eq.~\eqref{eq.Lag.eff.SMEFT} generates a Majorana mass term for neutrinos:
\begin{equation}
\mathcal{L}_{\Delta L = 2} = -\frac{1}{2} (M_\nu)_{\ell\ell^{\prime}} \, \nu^T_{\ell L} C \nu_{\ell^{\prime} L} + \text{h.c.}\ ,\quad (M_\nu)_{\ell\ell^{\prime}} = y_{\ell\ell^{\prime}} \frac{v^2}{\Lambda_{\rm LNV}}\ .
\end{equation}
The mass matrix $M_\nu$ is symmetric but non-diagonal. It can be diagonalized using the neutrino mass eigenstates $\nu_{iL}$ ($i=1,2,3$). The neutrino flavor eigenstates $\nu_{\ell L}$ ($\ell=e,\mu,\tau$) are related to the mass eigenstates by the Pontecorvo-Maki-Nakagawa-Sakata (PMNS) mixing matrix $U$ via
$\nu_{\ell L} = \sum_{i=1}^3 U_{\ell i} \nu_{iL}$. 
For the hyperon $0\nu\beta\beta$ decays considered below, we only need the diagonal element $(M_\nu)_{\ell\ell}$ relevant to the $\ell^-\ell^-$ final state,  
and the LNV Lagrangian is recast into
\begin{align}\label{eq.lag.DeltaL2}
\mathcal{L}_{\Delta L = 2} = -\frac{1}{2} m_{\ell\ell} \, \nu^T_{\ell L} C \nu_{\ell L} + \text{h.c.} \ ,
\end{align}
where $m_{\ell\ell}  \equiv \left(M_\nu\right)_{\ell\ell} =  \sum_{i=1}^3 U_{\ell i}^2 \, m_{i}$ is the so-called effective Majorana mass for the neutrino flavor eigenstates, with $m_i$ being the mass of the $\nu_{iL}$ state.

For the energy scale $\Lambda_{\chi}<\mu\ll\Lambda_{\rm EM}$, where $\Lambda_\chi\sim 1~{\rm GeV}$ represents the chiral symmetry breaking scale, the heavy degrees of freedom, the Higgs and weak gauge bosons, can be integrated out. By using the external source method, the low-energy effective Lagrangian can be written as
\begin{align}
\mathcal{L}_{\rm eff}^{\rm LE} =\mathcal{L}_{\rm QCD}^0 + \big[\bar{q}_L\gamma^\mu l_\mu q_L+ \bar{q}_R\gamma^\mu r_\mu q_R+\bar{q}(s-{\rm i}\gamma_5p)q\big]+\mathcal{L}_{\Delta L=2}+\cdots\ ,
\label{eq:Lagrangian}
\end{align}
where $\mathcal{L}_{\rm QCD}^0$ is the QCD Lagrangian in the SU(3) chiral limit and exhibits a global chiral ${\rm SU(3)}_L \times {\rm SU(3)}_R$ symmetry; the dots represent terms irrelevant to our calculation. In eq.~\eqref{eq:Lagrangian}, $s,p,l_{\mu},r_{\mu}$ are the external scalar, pseudoscalar, and left- and right-handed vector sources, respectively. The light quark mass matrix can be introduced through 
\begin{align}\label{eq.pands}
p=0\ ,\quad s={\rm diag}(m_u,m_d,m_s)\ ,
\end{align}
which characterizes the effect of explicit chiral symmetry breaking. 
Note that the scalar source $s$ in eqs.~\eqref{eq:Lagrangian} and~\eqref{eq.pands} is unrelated to the Mandelstam variable $s$ used for kinematics in section~\ref{sec:kinematics}.
The charged current weak interaction among quarks, neutrinos $\nu_{\ell L}=(\nu_{eL},\nu_{\mu L},\nu_{\tau L})$ and leptons $\ell_{L}=(e_L,\mu_L,\tau_L)^T$ can be incorporated by identifying the left- and right-handed external sources as
\begin{align}\label{eq.landr}
l_{\mu}&=-2\sqrt{2}G_FT_+[\bar{\nu}_{\ell L}\gamma_\mu\ell_L]+{\rm h.c.}\ , \quad	r_{\mu}=0\ ,
\end{align}
where $G_F$ is the Fermi constant and
\begin{align}\label{eq:T}
T_+=\left(\begin{array}{ccc}
   0  & V_{ud} & V_{us}\\
   0  & 0 & 0 \\
   0  & 0 & 0
\end{array}\right)=V_{ud} \hat{T}_++ V_{us}\hat{V}_+\ .
\end{align}
Here, $\hat{T}_+=(\lambda_1+{\rm i}\lambda_2)/2$ and $\hat{V}_+=(\lambda_4+{\rm i}\lambda_5)/2$~\cite{Thomson:2013zua} are ladder operators with $\lambda^a$ ($a=1,\cdots,8$) being the standard Gell-Mann matrices in flavor space. 
They change the $d$-type quarks to $u$ types, i.e., $d\to u$ and $s\to u$, respectively. 
$V_{ij}$ denotes the elements of the Cabibbo-Kobayashi-Maskawa (CKM) quark-mixing matrix. 

For the energy scale $\mu <\Lambda_\chi$, the chiral ${\rm SU(3)}_L\times {\rm SU(3)}_R$ symmetry is spontaneously broken down to the SU(3)$_V$ group, accompanied by the emergence of Goldstone bosons. Eventually, the chiral effective Lagrangian with hadrons as explicit degrees of freedom can be deduced as
\begin{align}\label{eq.lag.eff}
    \mathcal{L}_{\rm eff}^\chi = \mathcal{L}_{M}^{(2)} + \mathcal{L}_{MB}^{(1)} + \mathcal{L}_{\Delta L=2} +  \cdots,
\end{align}
where the superscripts indicate the chiral orders, the dots denote terms involving higher order operators, and $\mathcal{L}_{\Delta L=2}$ inherits eq.~\eqref{eq.lag.DeltaL2}. The first term $\mathcal{L}_{M}^{(2)}$ describes pure mesonic interaction at leading order (LO), which reads~\cite{Gasser:1983yg}
\begin{equation}
	\mathcal{L}_{M}^{(2)}=\frac{F_0^2}{4}{\rm Tr}[(D_{\mu}U)^{\dagger}D^{\mu}U]+\frac{F^2_0}{4}{\rm Tr}[U^{\dagger}\chi+U\chi^{\dagger}]\ ,
    \label{eq:LagM}
\end{equation}
where the SU(3) matrix $U$ represents the nonlinear realization of the Nambu-Goldstone fields through the exponential parametrization
\begin{align}\label{eq:Baryon_matrix}
 U =\exp\bigg\{\frac{{\rm i}\phi^a\lambda^a}{F_0}\bigg\}\ ,\quad  \sum_{a=1}^8\frac{\phi^a\lambda^a}{\sqrt{2}} = \left(\begin{array}{ccc}
\frac{1}{\sqrt{2}}\pi^0+\frac{1}{\sqrt{6}}\eta & \pi^+ & K^+\\
\pi^-&-\frac{1}{\sqrt{2}}\pi^0+\frac{1}{\sqrt{6}}\eta& K^0\\
K^-&\bar{K}^0&-\frac{2}{\sqrt{6}}\eta
\end{array}\right)\ ,
\end{align}
with $F_0$ the Goldstone-boson decay constant in the SU(3) chiral limit. The covariant derivative $D_\mu$ and the building block $\chi$ are given by
\begin{equation}
	D_{\mu}U=\partial_{\mu}U-{\rm i}l_{\mu}U+{\rm i}Ur_{\mu}\ ,	\quad \chi=2B_0(s+{\rm i}p)\ .
\end{equation}
The constant $B_0$ is proportional to the quark condensate in the chiral limit, and the external sources $s$, $p$, $l_\mu$ as well as $r_\mu$ are assigned according to eqs.~\eqref{eq.pands} and~\eqref{eq.landr}.

The second term of eq.~\eqref{eq.lag.eff} corresponds to the LO interaction between the pseudoscalar mesons and the octet baryons~\cite{Georgi:1984zwz,Krause:1990xc,Oller:2006yh}
\begin{equation}
	\mathcal{L}^{(1)}_{M B}={\rm Tr}\left[\bar{B}\left({\rm i} \slashed{D}-m\right)B\right]
	-\frac{D}{2}\left\langle \bar{B}\gamma^\mu\gamma_5\left\{u_\mu, B\right\}\right\rangle
	-\frac{F}{2}\left\langle \bar{B}\gamma^\mu\gamma_5\left[u_\mu, B\right]\right\rangle\ ,
\end{equation}
where $m$ is the baryon mass in the chiral limit, and $D$ and $F$ are parameters that can be determined by semileptonic baryon decays. Here, the axial vielbein $u_{\mu}$ is given by
\begin{align}
	u_{\mu}={\rm i}[u^{\dagger}(\partial_{\mu}-{\rm i}r_{\mu})u-u(\partial_{\mu}-{\rm i}l_{\mu})u^{\dagger}]\ , \quad u=\sqrt{U}\ .
\end{align}
The octet baryon fields are collected in the traceless matrix $B$ as
\begin{align}
    B=\sum_{a=1}^8\frac{B_a\lambda_a}{\sqrt{2}}=
\left(\begin{array}{ccc}
\frac{1}{\sqrt{2}}\Sigma^0+\frac{1}{\sqrt{6}}\Lambda & \Sigma^+ & p\\
\Sigma^-&-\frac{1}{\sqrt{2}}\Sigma^0+\frac{1}{\sqrt{6}}\Lambda&n\\
\Xi^-&\Xi^0&-\frac{2}{\sqrt{6}}\Lambda
\end{array}\right)\ .
\end{align}
The covariant derivative acting on the baryon fields is defined as
\begin{equation}
D_{\mu}B=\partial_{\mu}B+[\Gamma_{\mu},B]\ ,\quad \Gamma_{\mu}=\frac{1}{2}[u^{\dagger}(\partial_{\mu}-{\rm i}r_{\mu})u+u(\partial_{\mu}-{\rm i}l_{\mu})u^{\dagger}]\ .
\end{equation}

\subsection{Anatomy of the amplitude}
\label{sec:Amp}

\begin{figure}[ht]
\centering
\includegraphics[width=0.4\linewidth]{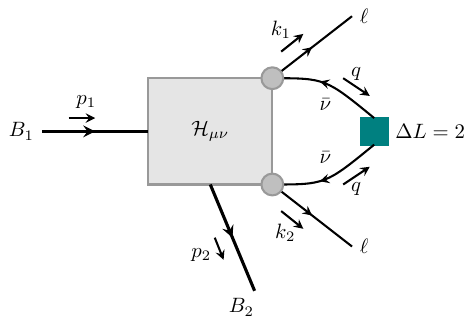}
\caption{Mass mechanism of hyperon $0\nu\beta\beta$ decays.}
\label{fig.mass.mech}
\end{figure}

The exchange of light Majorana neutrinos is the standard and most widely studied mechanism for the $0\nu\beta\beta$ decay; see e.g., ref.~\cite{DellOro:2016tmg} for review. The mass mechanism of $0\nu\beta\beta$ decays of hyperons is displayed in figure~\ref{fig.mass.mech}.  The teal boxes indicate the LNV vertices with $\Delta L=2$, while the gray filled circles correspond to the SM charged-current vertices. The big gray filled box stands for the hadronic part $\mathcal{H}_{\mu\nu}$, comprising the strong interactions between the Goldstone bosons and baryons.

The decay amplitude $\mathcal{M}_{0\nu\beta\beta}$, responsible for the long-range light-neutrino exchange mechanism, can be formally written as 
\begin{equation}
\label{eq:M}
	\mathcal{M}_{0\nu\beta\beta}=\int\frac{{\rm d}^d q}{(2\pi)^d}\mathcal{H}_{\mu\nu}(p_1,p_2;q)\mathcal{L}^{\mu\nu}(k_1,k_2;q)\ ,
\end{equation}
where we denote the leptonic part by $\mathcal{L}^{\mu\nu}$. From figure~\ref{fig.mass.mech}, one can read off the explicit expression for $\mathcal{L}^{\mu\nu}$:  
\begin{align}
  \mathcal{L}^{\mu\nu}(k_1,k_2;q)  =&\,\sum_{s,s^
  \prime}\big[-2\sqrt{2}G_F \bar{u}_{\ell\,L}(k_1)\gamma^{\mu} \upsilon^{(s)}_{\ell\,L}(q)\big]\frac{\rm i}{q^2+{\rm i}\epsilon}\big[\bar{\upsilon}^{(s)}_{\ell\,L}(q)(-\frac{\rm i}{2}m_{\ell\ell}C)\bar{\upsilon}^{(s^\prime),T}_{\ell\,L}(-q)\big]\notag\\
  &\times\frac{\rm i}{(-q)^2+{\rm i}\epsilon}\big[-2\sqrt{2}G_F \bar{u}_{\ell\,L}(k_2)\gamma^{\nu} \upsilon^{(s^\prime)}_{\ell\,L}(-q)\big]\ .\label{eq.leptonic.tensor.original}
\end{align}
The summation runs over the spins $s,s^\prime$ of the intermediate neutrinos.
As demonstrated in appendix~\ref{app:lept}, it can be simplified into the following form
\begin{align}
    \mathcal{L}^{\mu\nu}(k_1,k_2;q)=C_{\rm Lept}\cdot S (q^2)\cdot L^{\mu\nu}(k_1,k_2)\ ,\label{eq.leptonic.part.reduced}
\end{align}
where the leptonic coefficient $C_{\rm Lept}$ and Majorana neutrino propagator are defined by
\begin{align} 
    C_{\rm Lept} &= 4m_{\ell\ell}G_F^2V_{ui}V_{uj}\ ,\label{eq.coe.lept} \\
    S(q^2)& = \frac{\rm i}{q^2+{\rm i}\epsilon}\ ,
\end{align}
respectively. Here $V_{ui}$ and $V_{uj}$ ($i,j\in\{d,s\}$) are CKM matrix elements. The explicit expression of the last term in eq.~\eqref{eq.leptonic.part.reduced} reads
\begin{align} 
\label{eq:Lmunu}
   L^{\mu\nu}(k_1,k_2)&=\bar{u}_{\ell\,L}(k_1)\gamma^{\mu}\gamma^{\nu} C \bar{u}_{\ell\,L}^T(k_2)\ ,
\end{align}
which is referred to as the leptonic tensor hereafter.
Substituting eq.~\eqref{eq.leptonic.part.reduced} into eq.~\eqref{eq:M} and factoring out the $C_{\rm Lept}$ and $L^{\mu\nu}(k_1,k_2)$ from the loop integral, the amplitude $\mathcal{M}_{0\nu\beta\beta}$ can be recast as
\begin{align}\label{eq.amplitude.decomposition}
  \mathcal{M}_{0\nu\beta\beta}=  C_{\rm Lept}\, {H}_{\mu\nu}(p_1,p_2;k_1,k_2)\, L^{\mu\nu}(k_1,k_2) ,
\end{align}
with the hadronic tensor given by
\begin{align}\label{eq.amplitude.hadronic}
  {H}_{\mu\nu}(p_1,p_2;k_1,k_2)=  \int\frac{{\rm d}^d q}{(2\pi)^d}\mathcal{H}_{\mu\nu}(p_1,p_2;q) S(q^2)\ .
\end{align}
The hadronic tensor can be further decomposed as
\begin{align}\label{eq.lorentz.decomposition}
&H_{\mu\nu}(s,t,u)=\bar{u}(p_2)\sum_{i=1}^{34}\big[{V}_i(s,t,u)\mathcal{O}_{V,\mu\nu}^i+{A}_i(s,t,u)\mathcal{O}_{A,\mu\nu}^{i}\big]u(p_1)\ ,
\end{align}
where $\mathcal{O}^i_{\mu\nu}$ and $\mathcal{O}_{\mu\nu}^{Ai}$ denote the parity-even and parity-odd Lorentz operators, respectively. Explicit forms of the parity-even operators are given by
\begin{align}
\begin{array}{llll}
\label{tab:operators}
\mathcal{O}^1_{V,\mu\nu}=g_{\mu\nu}\ ,
&\mathcal{O}^2_{V,\mu\nu}=\slashed{k}_1 g_{\mu\nu}\ ,
&\mathcal{O}^3_{V,\mu\nu}=\gamma_{\mu}\gamma_{\nu}\slashed{k}_1\ ,
&\mathcal{O}^4_{V,\mu\nu}=\gamma_{\mu}\gamma_{\nu}\ ,\\
\mathcal{O}^5_{V,\mu\nu}=k_{1\mu}k_{2\nu}\ ,
&\mathcal{O}^6_{V,\mu\nu}=k_{1\nu}k_{2\mu}\ ,
&\mathcal{O}^7_{V,\mu\nu}=k_{1\mu}k_{1\nu}\ ,
&\mathcal{O}^8_{V,\mu\nu}=k_{2\mu}k_{2\nu}\ ,\\
\mathcal{O}^9_{V,\mu\nu}=p_{1\mu}p_{1\nu}\ ,
&\mathcal{O}^{10}_{V,\mu\nu}=p_{1\mu}k_{1\nu}\ ,
&\mathcal{O}^{11}_{V,\mu\nu}=p_{1\mu}k_{2\nu}\ ,
&\mathcal{O}^{12}_{V,\mu\nu}=p_{1\nu}k_{1\mu}\ ,\\
\mathcal{O}^{13}_{V,\mu\nu}=p_{1\nu}k_{2\mu}\ ,\quad
&\mathcal{O}^{14}_{V,\mu\nu}=k_{1\mu}k_{2\nu}\slashed{k}_1\ ,
&\mathcal{O}^{15}_{V,\mu\nu}=k_{1\nu}k_{2\mu}\slashed{k}_1\ ,
&\mathcal{O}^{16}_{V,\mu\nu}=k_{1\mu}k_{1\nu}\slashed{k}_1\ ,\\
\mathcal{O}^{17}_{V,\mu\nu}=k_{2\mu}k_{2\nu}\slashed{k}_1\ ,
&\mathcal{O}^{18}_{V,\mu\nu}=p_{1\mu}p_{1\nu}\slashed{k}_1\ ,
&\mathcal{O}^{19}_{V,\mu\nu}=p_{1\mu}k_{1\nu}\slashed{k}_1\ ,
&\mathcal{O}^{20}_{V,\mu\nu}=p_{1\mu}k_{2\nu}\slashed{k}_1\ ,\\
\mathcal{O}^{21}_{V,\mu\nu}=p_{1\nu}k_{1\mu}\slashed{k}_1\ ,
&\mathcal{O}^{22}_{V,\mu\nu}=p_{1\nu}k_{2\mu}\slashed{k}_1\ ,
&\mathcal{O}^{23}_{V,\mu\nu}=p_{1\mu}\gamma_{\nu}\ ,
&\mathcal{O}^{24}_{V,\mu\nu}=p_{1\nu}\gamma_{\mu}\ ,\\
\mathcal{O}^{25}_{V,\mu\nu}=k_{1\mu}\gamma_{\nu}\ ,
&\mathcal{O}^{26}_{V,\mu\nu}=k_{2\mu}\gamma_{\nu}\ ,
&\mathcal{O}^{27}_{V,\mu\nu}=k_{1\nu}\gamma_{\mu}\ ,
&\mathcal{O}^{28}_{V,\mu\nu}=k_{2\nu}\gamma_{\mu}\ ,\\
\mathcal{O}^{29}_{V,\mu\nu}=p_{1\mu}\gamma_{\nu}\slashed{k}_1\ ,
&\mathcal{O}^{30}_{V,\mu\nu}=p_{1\nu}\gamma_{\mu}\slashed{k}_1\ ,
&\mathcal{O}^{31}_{V,\mu\nu}=k_{1\mu}\gamma_{\nu}\slashed{k}_1\ ,
&\mathcal{O}^{32}_{V,\mu\nu}=k_{2\nu}\gamma_{\mu}\slashed{k}_1\ ,\\
\mathcal{O}^{33}_{V,\mu\nu}=k_{2\mu}\gamma_{\nu}\slashed{k}_1\ ,
&\mathcal{O}^{34}_{V,\mu\nu}=k_{1\nu}\gamma_{\mu}\slashed{k}_1\ ,
\end{array}
\end{align}
while the parity-odd ones are defined via
\begin{align}\label{tab:operators_OA}
    \mathcal{O}_{A,\mu\nu}^i = \mathcal{O}_{V,\mu\nu}^i\gamma_5\ ,\quad i=1,\cdots,34\ .
\end{align}
The above set of operators is complete but not minimal. The corresponding coefficients $V_i(s,t,u)$ and $A_i(s,t,u)$ are structure functions of Mandelstam variables $s$, $t$, and $u$.
They encode the information of underlying strong dynamics and can be computed from the Feynman diagrams shown in figure~\ref{fig.Feynman.Diagram}. 

\subsection{One-loop decay amplitudes}
\label{sec:OneLoop_Amp}
\begin{figure}[ht]
\centering
\includegraphics[width=0.9\linewidth]{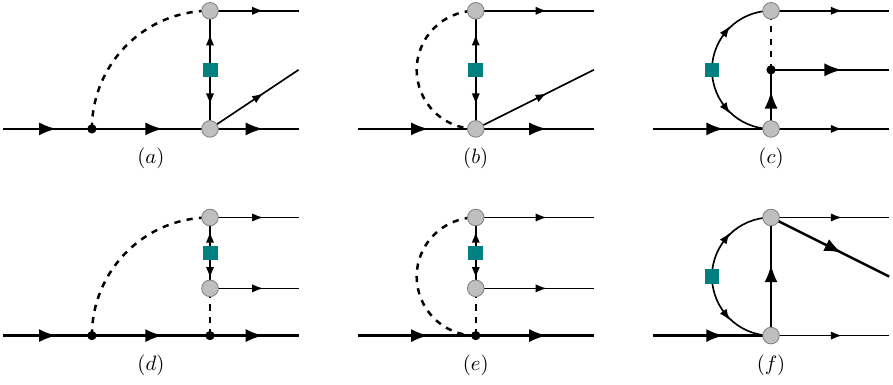}
\caption{One-loop Feynman diagrams contributing to the $0\nu \beta\beta$ decay of hyperons. The thick solid, thin solid and dashed lines represent baryons, leptons and pions, in order. The teal box denotes the LNV vertex with $\Delta L= 2$, while the gray filled circles stand for charged-current vertices from the SM. The crossed diagrams are not shown explicitly.}
\label{fig.Feynman.Diagram}
\end{figure}

The one-loop Feynman diagrams relevant to our calculation are displayed in figure~\ref{fig.Feynman.Diagram}. Their crossed partners, obtained by interchanging the final leptons, are not shown explicitly. Note that diagram (f) was computed in a model-dependent way in ref.~\cite{Barbero:2007zm}. It can be seen that the one-loop diagrams exhibit the same structure as the mass mechanism shown in figure~\ref{fig.mass.mech}. Therefore, the master formulae~\eqref{eq.amplitude.decomposition} and~\eqref{eq.amplitude.hadronic}, derived in the preceding subsection, can be readily applied to formulate the loop amplitudes.
In the same form as eq.~\eqref{eq.amplitude.hadronic}, the hadronic tensors for each one-loop diagram are listed in appendix~\ref{app:hadronic_Amp}. The loop integrations can be performed straightforwardly by using, e.g., \texttt{FeynCalc}~\cite{Mertig:1990an, Shtabovenko:2016sxi, Shtabovenko:2020gxv}. We have obtained all the pertinent hadronic structure functions $V_i$ and $A_i$ ($i=1,\cdots,34$) for the four physical $0\nu\beta\beta$ decay processes in eq.~\eqref{eq.physical.processes}. However, the resultant expressions are too lengthy to be explicitly shown here. 

In what follows, we illustrate how to derive the amplitudes for the crossed diagrams. For clarity, we refer to the diagrams in figure~\ref{fig.Feynman.Diagram} as the $u$-channel contributions. When the momenta of the two final-state leptons are interchanged while keeping the Lorentz indices at the charged-current vertices fixed, one obtains the $t$-channel contributions from the crossed diagrams. The $t$- and $u$-channel amplitudes satisfy the relation 
\begin{align}
    \mathcal{M}_{0\nu\beta\beta}^t = - \mathcal{M}_{0\nu\beta\beta}^u (k_1\leftrightarrow k_2, u \leftrightarrow t)\ ,
\end{align}
where the minus sign arises from the interchange of the two identical leptons in the final states, as required by the Fermi-Dirac statistics. More specifically, one has
\begin{align}
    \mathcal{M}_{0\nu\beta\beta}^t&=  -C_{\rm Lept}\, {H}^u_{\mu\nu}(p_1,p_2;k_2,k_1)\, L^{\mu\nu}(k_2,k_1)\notag \\
     &=C_{\rm Lept}\, {H}^u_{\mu\nu}(p_1,p_2;k_2,k_1)\, L^{\nu\mu}(k_1,k_2)\notag \\
     &=C_{\rm Lept}\, {H}^u_{\nu\mu}(p_1,p_2;k_2,k_1)\, L^{\mu\nu}(k_1,k_2)\ .
\end{align}
Note that the second equality holds due to eq.~\eqref{eq:su}, whereas the last equality is established by relabeling the dummy indices $\mu$ and $\nu$. Therefore, the $t$- and $u$-channel hadronic tensors are related to each other by
\begin{align}\label{eq.tu.hadronic.tensor}
  {H}^t_{\mu\nu}(p_1,p_2;k_1,k_2) = {H}^u_{\nu\mu}(p_1,p_2;k_2,k_1)\ .
\end{align}
Applying Lorentz decomposition to both sides yields the relations between the $t$- and $u$-channel hadronic structure functions, which are summarized in appendix~\ref{app:cross}. Finally, the full hadronic tensor is given by
\begin{align}
  {H}_{\mu\nu}(p_1,p_2;k_1,k_2)  = {H}^u_{\nu\mu}(p_1,p_2;k_2,k_1)+{H}^u_{\mu\nu}(p_1,p_2;k_1,k_2)\ .\label{eq.hadron.tensor.total}
\end{align}

Before ending this subsection, it is worth discussing the chiral power counting of the Feynman diagrams and the decay amplitudes. We denote the small parameters in the chiral expansion collectively by $p$. By counting the light neutrino effective propagator as of order $p^{-2}$, the standard power counting rule applies to all the Feynman diagrams. In this counting rule, a four-dimensional loop integration is of $\mathcal{O}(p^4)$, an vertex obtained from $\mathcal{O}(p^n)$ Lagrangian counts as $\mathcal{O}(p^n)$, a Goldstone boson propagator as $\mathcal{O}(p^{-2})$ and a baryon propagator as $\mathcal{O}(p^{-1})$. To conclude, for a given diagram with $L$ loops, $V_n$ vertices of $\mathcal{O}(p^n)$, $I_M$ meson propagators  and $I_B$ baryon propagators, the chiral dimension is assigned to be
\begin{align}
    D_\chi = 4L+\sum_n n V_n- 2I_M-I_B-2 \ ,\label{eq.PCR}
\end{align}
where the last factor $-2$ is responsible for the Majorana neutrino propagator. 
Therefore, the one-loop diagrams in figure~\ref{fig.Feynman.Diagram} are of the same order, i.e., $\mathcal{O}(p^3)$. It is also worth mentioning that the leptonic tensor $L_{\mu\nu}$, stemming from two charged weak currents $l_\mu$ and $l_\nu$ of $\mathcal{O}(p)$ (c.f. eq.~\eqref{eq.landr}), should have the order of $p^2$. In consequence, the one-loop hadronic tensor $H_{\mu\nu}$ is $\mathcal{O}(p)$. 

For completeness, we also provide the chiral orders of the Lorentz operators involved in the hadronic tensor as follows:
\begin{align}
\mathcal{O}(1):&~~\mathcal{O}_V^{1,4,9,23,24}\sim  \mathcal{O}_A^{4,23,24} \ ,\notag\\
\mathcal{O}(p):&~~\mathcal{O}_V^{2,3,10-13,18,25-30}\sim  \mathcal{O}_A^{1-3,9,18,25-30} \ ,\notag \\
\mathcal{O}(p^2):&~~ \mathcal{O}_V^{5-8,19-22,31-34} \sim  \mathcal{O}_A^{10-13,19-22,31-34}\ ,\notag\\
\mathcal{O}(p^3):&~~ \mathcal{O}_V^{14-17} \sim  \mathcal{O}_A^{5-8,14-17} \ .
\end{align}
The Lorentz indices of the operators are suppressed henceforth if no confusion is caused. In addition, the masses and Mandelstam variables count as
\begin{align}
  m\sim \mathcal{O}(1)\ ,\quad (t-m^2)\sim (u-m^2)\sim \mathcal{O}(p)\ ,\quad  s\sim M_\pi^2\sim M_K^2\sim M_\eta^2 \sim \mathcal{O}(p^2)\ ,
\end{align}
where $m$ signifies the baryon mass, either in the SU(3) chiral limit or in the physical case.

\subsection{Renormalization}
\label{sec:Renormalization}
We employ DR to calculate the loop diagrams in $d$-dimensional space-time, and the UV divergence is extracted by using the $\overline{\rm MS}-1$ subtraction scheme. For the four $0\nu\beta\beta$ decays exhibited in eq.~\eqref{eq.physical.processes}, we find that the UV divergences of their hadronic tensors at one-loop order turn out to be of the same form as
\begin{align}
    H^{\rm UV}_{\mu\nu}=&\bigg[\frac{1}{2}m(2D^2-6F^2+1)\mathcal{O}_{V}^{1} + \frac{1}{6}(D^2-3F^2-3)\big(-\frac{3}{2}\mathcal{O}_{V}^2 + \mathcal{O}_{V}^3 - 2\mathcal{O}_{V}^{23}+3\mathcal{O}_{V}^{25} \notag\\
    & + \mathcal{O}_{V}^{26}\big) + F\big(m\mathcal{O}_{A}^1 + \frac{3}{2}\mathcal{O}_{A}^2 - \mathcal{O}_{A}^3 - m\mathcal{O}_{A}^4 + 2\mathcal{O}_{A}^{23} - 3\mathcal{O}_{A}^{25} - \mathcal{O}_{A}^{26}\big)\bigg]R\ ,
\end{align}
where $R=2/(d-4)+\gamma_E-1-\ln(4\pi)$ with $\gamma_E=0.5772$ the Euler's constant.

In addition to the UV divergence, a further subtlety of loop calculations in baryon ChPT is the emergence of the so-called PCB problem. This problem is caused by the non-vanishing baryon mass in the chiral limit. For instance, for a given loop diagram of $\mathcal{O}(p^{D_\chi})$ with $D_\chi$ specified by the naive power counting rule~\eqref{eq.PCR}, the resultant loop amplitude regulated in DR contains terms of orders lower than $D_\chi$. In our current case, the one-loop hadronic tensor $H_{\mu\nu}$ should be $\mathcal{O}(p)$, however, PCB terms of $\mathcal{O}(1)$ show up:
\begin{align}
    H^{\rm PCB}_{\mu\nu}=&\bigg[\frac{-(D^2-3F^2)(d^2+5d-10)+3d(3-d)}{6d(d-3)m}\mathcal{O}_{V}^{1} +\frac{2D^2-6F^2-3}{3(d-3)m^2}\mathcal{O}_{V}^{23}\notag\\
    & + \frac{F}{(d-3)m}\mathcal{O}_{A}^{4} - \frac{2F}{(d-3)m^2}\mathcal{O}_{A}^{23} \bigg]A_0(m^2) \ ,
\end{align}
where $A_0(m^2)$ denotes the standard one-point scalar integral in DR, as defined in eq.~\eqref{eq.ABCD}.

The emergence of UV and PCB terms necessitates the introduction of counterterms. Under the chiral ${\rm SU(3)}_L \times {\rm SU(3)}_R$ group, the meson and baryon fields transform as $u \rightarrow LuK^{\dagger}=KuR^{\dagger}, B \rightarrow KBK^{\dagger}, \bar{B} \rightarrow K\bar{B}K^{\dagger}$~\cite{Coleman:1969sm,Callan:1969sn}, respectively. Furthermore, $T^+$ can be regarded as a spurion transforming as $T^+ \rightarrow LT^+L^{\dagger}$. We construct the relevant operators invariant under the above specified chiral transformation, 
\begin{align}
\label{eq:counterterm}
    \mathcal{L}^{C}&=  \mathcal{L}_{1}^{C} +\mathcal{L}_{2}^{C} +\mathcal{L}_{3}^{C} \ ,\\
   \mathcal{L}_{1}^{C} 
   &=
  4m_{\ell\ell}G_F^2 \text{Tr}[\bar B u^\dagger T^+ u] g_{\mu\nu} (g_1 + g_1^\prime \gamma_5) \text{Tr}[u^\dagger T^+ u B] \bar\ell_L \gamma^\mu\gamma^\nu C \bar\ell^T_L + \rm{h.c.}\ ,\\
   \mathcal{L}_{2} ^{C}  
   &=
    4{\rm i}m_{\ell\ell}G_F^2\text{Tr}[\bar B u^\dagger T^+ u] \gamma_{\mu} \gamma_\nu \gamma_\rho(g_2 + g_2^\prime \gamma_5) \text{Tr}[u^\dagger T^+ u B] \bar\ell_L \gamma^\mu\gamma^\nu C (\partial^\rho \bar\ell^T_L) + \rm{h.c.}\ ,\\
   \mathcal{L}_{3} ^{C}  
   &=
    4{\rm i}m_{\ell\ell}G_F^2\text{Tr}[\bar B u^\dagger T^+ u] \gamma_{\mu} (g_3 + g_3^\prime \gamma_5) \text{Tr}[u^\dagger T^+ u B] \bar\ell_L \gamma^\mu\gamma^\nu C (\partial_\nu \bar\ell^T_L) + \rm{h.c.}\ .
\end{align}
In fact, the operators $\text{Tr}[\bar{B} u^\dagger T^+ u]$ and $\text{Tr}[u^\dagger T^+ u B]$, where the traces are taken in flavor space, are each chirally invariant. It is straightforward to derive the counterterms from the above Lagrangians, which can be expressed as
\begin{align}
    H^{C}_{1,\mu\nu}=&\, 
    2 g_1 \mathcal{O}_{V}^1 + 2g_1^{\prime}\mathcal{O}_{A}^1 \ ,\notag\\
    H_{2,\mu\nu}^{C}=&
    -2g_2 \big[-\frac{2m}{d}\mathcal{O}_{V}^1+2(\mathcal{O}_{V}^{23}-\mathcal{O}_{V}^{25}-\mathcal{O}_{V}^{26}) + (\mathcal{O}_{V}^2 - \mathcal{O}_{V}^3) \big]\notag \\
    &- 2g_2^\prime \big[-\frac{2m}{d}\mathcal{O}_{A}^1-m\mathcal{O}_{A}^4 + 2(\mathcal{O}_{A}^{23}-\mathcal{O}_{A}^{25}-\mathcal{O}_{A}^{26}) + (\mathcal{O}_{A}^2-\mathcal{O}_{A}^3) \big]\ ,\notag\\
    H_{3,\mu\nu}^{C}=&
    g_3(\frac{2}{d}\mathcal{O}_{V}^2-\mathcal{O}_{V}^{25} + \mathcal{O}_{V}^{26}) - g_3^{\prime}(-\frac{4m}{d}\mathcal{O}_{A}^{1}-\frac{2}{d}\mathcal{O}_{A}^{2}+\mathcal{O}_{A}^{25}-\mathcal{O}_{A}^{26})\ .\label{eq.counterterm}
\end{align}
The UV divergences can be readily absorbed by splitting the coupling constants $g_i^{(\prime)}$ ($i=1,2,3$) into finite and infinite pieces. That is
\begin{align}
    g_i^{(\prime)} = g_i^{(\prime)r} + \frac{\beta_{g_i^{(\prime)}}}{16\pi^2}R\ ,
\end{align}
where $g_i^{(\prime)r}$ are UV-renormalized parameters. The second term cancels the UV divergences from loops, leading to the UV-$\beta$ functions
\begin{align}
\beta_{g_1} &= -\frac{m(11D^2-33F^2+9)}{24}\ , \quad  \beta_{g_2} = -\frac{1}{12}(D^2-3F^2-3)\ , 
\\ 
\beta_{g_3} &= \frac{1}{6}(D^2-3F^2-3)\ ,\quad \beta_{g_1^{\prime}} = -\frac{Fm}{4}\ , \quad
\beta_{g_2^{\prime}} = \frac{F}{2}\ , \quad
\beta_{g_3^{\prime}} = -F\ . 
\end{align}
The PCB terms can also be absorbed by the counterterms. We perform a finite-shift renormalization using the EOMS scheme~\cite{Fuchs:2003qc}. The UV-renormalized parameters are further decomposed as follows
\begin{align}
g_i^{(\prime)r} = \tilde{g}_i^{(\prime)}+\frac{\tilde{\beta}_{g_i^{(\prime)}}}{16\pi^2}\ ,
\end{align}
and the PCB terms are exactly canceled by the second term with the EOMS-$\tilde{\beta}$ functions given by
\begin{align}
\tilde{\beta}_{g_1} &= \frac{(d-2)[3(d-1)+(D^2-3F^2)(7+d)]}{12d(d-3)m}A_0(m^2)\ , \quad \tilde{\beta}_{g_1^{\prime}} = 0\ ,\\
\tilde{\beta}_{g_2} &= \frac{(2D^2-6F^2-3)A_0(m^2)}{12(d-3)m^2}\ , \quad \tilde{\beta}_{g_2^{\prime}} = -\frac{FA_0(m^2)}{2(d-3)m^2}\ ,\quad \tilde{\beta}_{g_3} = 0\ ,\quad \tilde{\beta}_{g_3^{\prime}} = 0\ .
\end{align}
The loop function $A_0(m^2)$ is defined in eq.~\eqref{eq.loopint}. Note that the parameters $g_3^r$, $g_1^{\prime r}$ and $g_3^{\prime r}$ are untouched by the procedure of EOMS finite renormalization.

\section{Numerical results and discussions}
\label{sec:numer}

In this section, we present predictions for the decay widths and branching ratios based on the one-loop amplitudes for hyperon $0\nu\beta\beta$ decays derived within the EOMS scheme of BChPT. We then define the neutrinoless TFFs and derive their chiral expressions. The dependence of these form factors on the dilepton invariant mass squared $s$ and the light $u/d$-quark mass is also analyzed.

\subsection{Differential decay rate and branching ratio}
\label{sec:Brach_ratio}

Based on eq.~\eqref{eq.partial.decay.rate}, the differential decay rate with respect to the di-lepton invariant mass squared $s$ is obtained by integrating out $u$, yielding
\begin{align}
\frac{d\,\Gamma}{{\rm d}s} =\frac{1}{2!}\frac{1}{(2\pi)^3}\frac{1}{32m_{1}^3}\int_{u_-}^{u_+}\overline{|\mathcal{M}_{0\nu\beta\beta}|^2}\,{\rm d}u\ ,
\end{align}
where the integration limits are specified in eq.~\eqref{eq.u.limits} and the amplitude $\mathcal{M}_{0\nu\beta\beta}$ has been derived in the previous section. The symmetry factor $1/(2!)$ accounts for the indistinguishability of the two identical final-state leptons, while the overline denotes the spin average over the initial-state hyperon. The amplitude squared $|\mathcal{M}_{0\nu\beta\beta}|^2$ can be evaluated using Casimir's trick, which allows us to express the spin sums as traces over Dirac matrices:
\begin{align}
\overline{|\mathcal{M}_{0\nu\beta\beta}|^2}=\frac{1}{2}\sum_{\rm spins} \mathcal{M}_{0\nu\beta\beta} \mathcal{M}_{0\nu\beta\beta}^\dagger
=\frac{1}{2}|C_{\rm Lept}|^2\, \hat{L}\cdot \hat{H}\ , 
\end{align}
where $\hat{L}$ and $\hat{H}$ denote the leptonic and hadronic rank-$4$ tensors, respectively. Their expressions are given by
\begin{align}
\hat{L}^{\mu\nu\alpha\beta}&={\rm Tr}[\gamma^\mu\gamma^\nu\slashed{k}_2 P_L \gamma^\beta\gamma^\alpha \slashed{k}_1 P_R]\ ,\\
\hat{H}_{\mu\nu\alpha\beta}&={\rm Tr}[H_{\mu\nu}(\slashed{p}_1+m_{1})\widetilde{H}_{\alpha\beta}(\slashed{p}_2+m_{2})]\ .
\end{align}
with $P_L=(1-\gamma_5)/2, P_R=({1+\gamma_5})/{2}$ and $\widetilde{H}_{\alpha\beta}\equiv\gamma_0{H}_{\alpha\beta}^\dagger\gamma_0$. Eventually, the formula for the differential decay rate becomes
\begin{align}
\frac{{\rm d}\,\Gamma}{{\rm d}\sqrt{s}}  ={\frac{1}{2!}}\frac{1}{(2\pi)^3}\frac{|C_{\rm Lept}|^2}{128m_{1}^3\sqrt{s} }\int_{u_{-}}^{u_{+}} \hat{L}\cdot \hat{H} \,{\rm d}u\ .
\end{align}

\begin{table}[tb]
\caption{Physical processes of hyperon $0\nu\beta\beta$ decays and their leptonic coefficient $C_{\rm Lept}$~defined by eq. \eqref{eq.coe.lept}. The effective neutrino mass $m_{\ell\ell}$ should be taken as $m_{ee}$ for the  electronic mode and $m_{\mu\mu}$ for the muonic mode. }
\label{tab:Clept}
\centering
\renewcommand{\arraystretch}{1.2}
\renewcommand{\tabcolsep}{1.0pc}
\begin{tabular}{c|ccc}
\hline
&\multicolumn{1}{c}{$\Delta S=0$} & \multicolumn{1}{c}{$\Delta S=1$} & \multicolumn{1}{c}{$\Delta S=2$} \\
\hline
 & $\Sigma^- \rightarrow \Sigma^+ e^- e^-$ & $\Sigma^- \rightarrow p\, e^- e^-$ & $\Xi^- \rightarrow p\, e^- e^-$ \\
Process &  & $\Sigma^- \to  p~ \mu^-\mu^-$ & $\Xi^- \to p\, \mu^- \mu^-$ \\
& & $\Xi^- \rightarrow \Sigma^+ e^- e^-$  & \\ \hline
$C_{\rm Lept}$ &
$4m_{\ell\ell}G_F^2V^2_{ud}$	&	$4m_{\ell\ell}G_F^2V_{ud}V_{us}$	&	$4m_{\ell\ell}G_F^2V^2_{us}$\\
\hline
\end{tabular}
\end{table}

We now present predictions for the differential decay rates and branching ratios based on the one-loop amplitude evaluated with renormalization scale $\mu$ set to the SU(3) chiral-limit baryon mass, i.e., $\mu=m$. There are six kinematically allowed hyperon $0\nu\beta\beta$ decay processes, which are listed in table~\ref{tab:Clept} along with their corresponding leptonic coefficients $C_{\rm Lept}$. In our numerical calculations, we adopt the CKM matrix elements $|V_{ud}|=0.97367$ and $|V_{us}|=0.22431$, the Fermi constant $G_F=1.16636 \times 10^{-5}~\rm GeV^{-2}$~\cite{ParticleDataGroup:2024}, and the LECs $D=0.8$ and $F=0.5$~\cite{Borasoy:1998pe}. 
We note that the Goldstone boson decay constant $F_0$ does not appear in the final amplitude $\mathcal{M}_{0\nu\beta\beta}$. Furthermore, the physical baryon and meson masses, taken from ref.~\cite{ParticleDataGroup:2024} and compiled in table~\ref{tab:octet_masses}, are employed in the phase space integration.
\begin{table}[tb]
\centering
\caption{Values of physical masses of the octet baryons and Goldstone bosons~\cite{ParticleDataGroup:2024}.}
\begin{tabular}{cc|cc} 
\hline
Particle & Mass (MeV) & Particle & Mass (MeV) \\
\hline
${p}$        & 938.3 & ${\Lambda}$ & 1111.5 \\
${n}$        & 939.6 & ${\Xi^0}$   & 1314.8 \\
${\Sigma^-}$ & 1197.5 & ${\Xi^-}$   & 1321.7 \\
${\Sigma^0}$ & 1192.6 & ${\pi^\pm}$     & 139.57 \\
${\Sigma^+}$ & 1189.4 & $K^\pm$         & 493.68 \\
\hline
\end{tabular}
\label{tab:octet_masses}
\end{table}

\begin{figure}[ht]
\centering
\includegraphics[width=\linewidth]{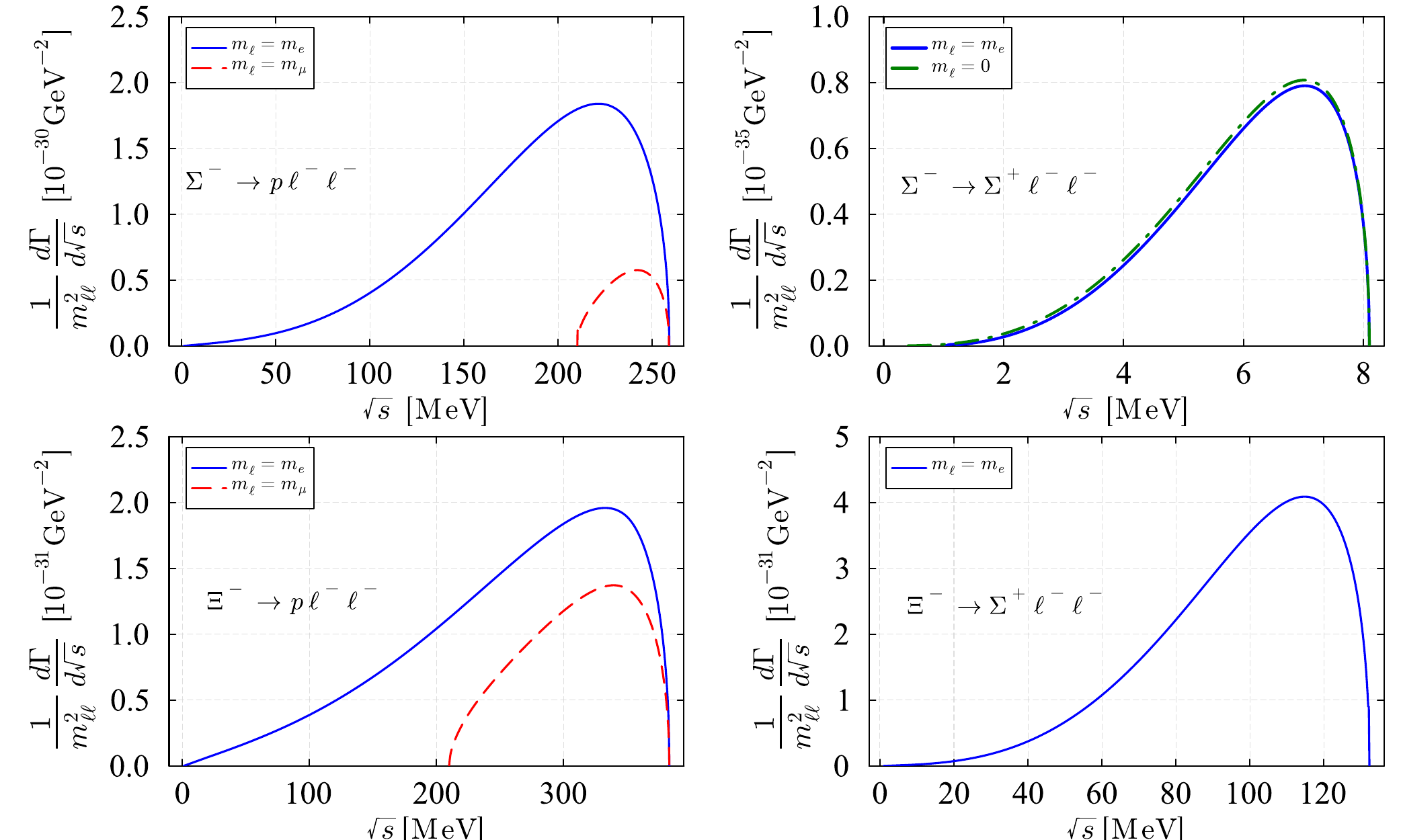}
\caption{Normalized differential decay distributions $\mathrm{d}\Gamma/\mathrm{d}\sqrt{s} \times (1/m_{\ell\ell}^2)$ for the LNV hyperon decay channels: $\Sigma^- \to p\, \ell^- \ell^-$ (top left), $\Sigma^- \to \Sigma^+\, \ell^- \ell^-$ (top right), $\Xi^- \to p\, \ell^- \ell^-$ (bottom left), and $\Xi^- \to \Sigma^+\, \ell^- \ell^-$  (bottom right). The blue solid and red dashed curves correspond to electronic and muonic modes, respectively. For comparison, the green dashed curve shows the decay rate in the massless lepton limit.}
\label{fig:LNV_decays}
\end{figure}

Our numerical results for the differential decay rates are displayed in figure~\ref{fig:LNV_decays}. We refer the reader to ref.~\cite{Zhao:2026wvf} for the  corresponding Dalitz plots of the squared amplitudes. They have been normalized by the effective neutrino mass squared $m_{\ell\ell}^2$, which contributes only as an overall factor.
Since the decay thresholds, defined as the mass difference between the initial and final baryons, are greater than $2m_\mu$ for the $(\Sigma^-,\Xi^-)\to p$ transitions, the final leptons can be a dielectron or a dimuon pair. In figure~\ref{fig:LNV_decays}, the decay rates for electronic and muonic decay modes are represented by blue solid and red dashed lines, respectively. Our results reveal that the primary contributions to the differential decay rates, across all channels, arise predominantly from the proximity of the decay threshold. Furthermore, the curves for $m_\ell=m_e$ and $m_\ell=0$ nearly overlap, as seen for the $\Sigma^-\to \Sigma^+$ decay in the top-right panel of figure~\ref{fig:LNV_decays}, implying that the effect of electron mass is negligible.

\begin{table}[ht]
\centering
\caption{Decay rates (normalized to the effective neutrino mass squared $m_{\ell\ell}^2$) and branching ratios for $\Delta L = 2$ hyperon decays, obtained from the one-loop amplitudes in the EOMS scheme with the renormalization scale $\mu=m$. The electron-mode branching ratios are evaluated using the benchmark value $m_{ee}^2= (100\, \text{meV})^2$, which is compatible with current nuclear $0\nu\beta\beta$ constraints under the standard light-neutrino mass mechanism. This value is nuclear-matrix-element dependent (arising from the translation of nuclear half-life limits) and should not be interpreted as a directly measured quantity. We use $m_{\mu \mu}^2 = (10\, \text{eV})^2$ to evaluate the muon-mode branching ratios.
}
\label{branch_ratio}
\setlength{\tabcolsep}{8pt} 
\begin{tabular}{c | c c c}
\hline
\multirow{2}{*}{Process} & \multirow{2}{*}{$\dfrac{\Gamma_{0\nu}}{m_{\ell\ell}^2}$ [sec$^{-1}$/MeV$^2$] }
 & \multicolumn{2}{c}{\makecell{$\mathcal{B}(B_1 \rightarrow B_2 \ell^- \ell^-)$}} \\
\cline{3-4} &  & \makecell{This work} & \makecell{Experiments} \\
\hline
$\Sigma^-\rightarrow p~ e^- e^-$ & $3.194 \times 10^{-7}$ & $4.7 \times 10^{-31}$ & $<6.7\times 10^{-5}$~\cite{BESIII:2020iwk}\\
$\Sigma^-\rightarrow \Sigma^+ e^- e^-$ & $3.925 \times 10^{-14}$ & $5.8 \times 10^{-38}$ & - \\
$\Sigma^-\rightarrow p~ \mu^- \mu^-$ & $3.202 \times 10^{-8}$ & $4.7 \times 10^{-28}$ & -\\
$\Xi^-\rightarrow \Sigma^+ e^- e^-$ & $3.404 \times 10^{-8}$ & $5.6 \times 10^{-32}$  &$<2.0\times 10^{-5}$~\cite{BESIII:2025ylz}\\
$\Xi^-\rightarrow p~e^- e^-$ & $5.706 \times 10^{-8}$ & $9.4 \times 10^{-32}$ & - \\
$\Xi^-\rightarrow p~\mu^- \mu^- $ & $2.509 \times 10^{-8}$ & $4.1 \times 10^{-28}$  & $<4.0\times 10^{-8}$~\cite{HyperCP:2005sby}\\
\hline
\end{tabular}
\end{table}

Our long-range one-loop results for partial decay widths and branching ratios can be found in the second and third columns of table~\ref{branch_ratio}, respectively. The partial decay widths are normalized by $m_{\ell\ell}^2$, so that they are solely determined by the phase-space factors and the corresponding chiral amplitudes without any unknown parameters. In the computation of the branching ratios, we use the lifetimes $\tau_{\Sigma^-}=1.479\times 10^{-10}~s$ for the $\Sigma^-$ and $\tau_{\Xi^-}=1.639\times10^{-10}~s$ for the $\Xi^-$~\cite{ParticleDataGroup:2024}. In addition, the effective electronic neutrino mass is chosen to be $m_{ee}=100$~meV,
where this benchmark value is chosen as a representative current bound, $m_{ee}=\mathcal{O}(0.1~{\rm eV})$, consistent with recent nuclear $0\nu\beta\beta$ limits~\cite{KamLAND-Zen:2024eml,CUORE:2024ikf,Majorana:2022udl,GERDA:2020xhi,EXO-200:2019rkq} within the standard light-neutrino mass mechanism, and should be understood as nuclear-matrix-element dependent. For the muonic mode, we take $ m_{\mu\mu}= 10\, \text{eV}$ as an illustrative benchmark for the effective Majorana mass, noting that the branching ratios scale as $m_{\ell\ell}^2$. It can be seen from table~\ref{branch_ratio} that the branching ratio of $\Sigma^-\to\Sigma^+e^-e^-$ is smaller than those of other channels by at least six orders of magnitude, owing to its tiny kinematic phase space. In fact, it is kinematically forbidden in the isospin limit. The experimental upper limits for the three decay processes $\Sigma^-\to pe^-e^-$~\cite{BESIII:2020iwk}, $\Xi^-\to pe^-e^-$~\cite{BESIII:2025ylz} and $\Xi^-\to p\mu^-\mu^-$~\cite{HyperCP:2005sby}, are also shown for easy comparison in table~\ref{branch_ratio}. 
Our results predicted by light-neutrino exchange are more than 20 orders of magnitude below the experimental upper bounds.
However, these predictions are made by just using the $\mathcal{O}(p^3)$ one-loop amplitudes but dropping the $\mathcal{O}(p^2)$ counterterms~\eqref{eq.counterterm}. These counterterms, being at a lower order than the one-loop amplitudes and corresponding to short-range contribution, should be more important than the one loops. In fact, the same finding was also observed in $nn\to pp e^-e^-$~\cite{Cirigliano:2018hja}. Inclusion of the $\mathcal{O}(p^2)$ counterterm contribution might enhance the branching ratios; however, one does not expect it to sizeably enhance the orders of magnitude of the branching ratios.  
Thus, this huge gap indicates that any nonvanishing observation would require mechanisms beyond light-neutrino exchange.

\begin{figure}[tb]
\centering
\includegraphics[width=0.7\textwidth]{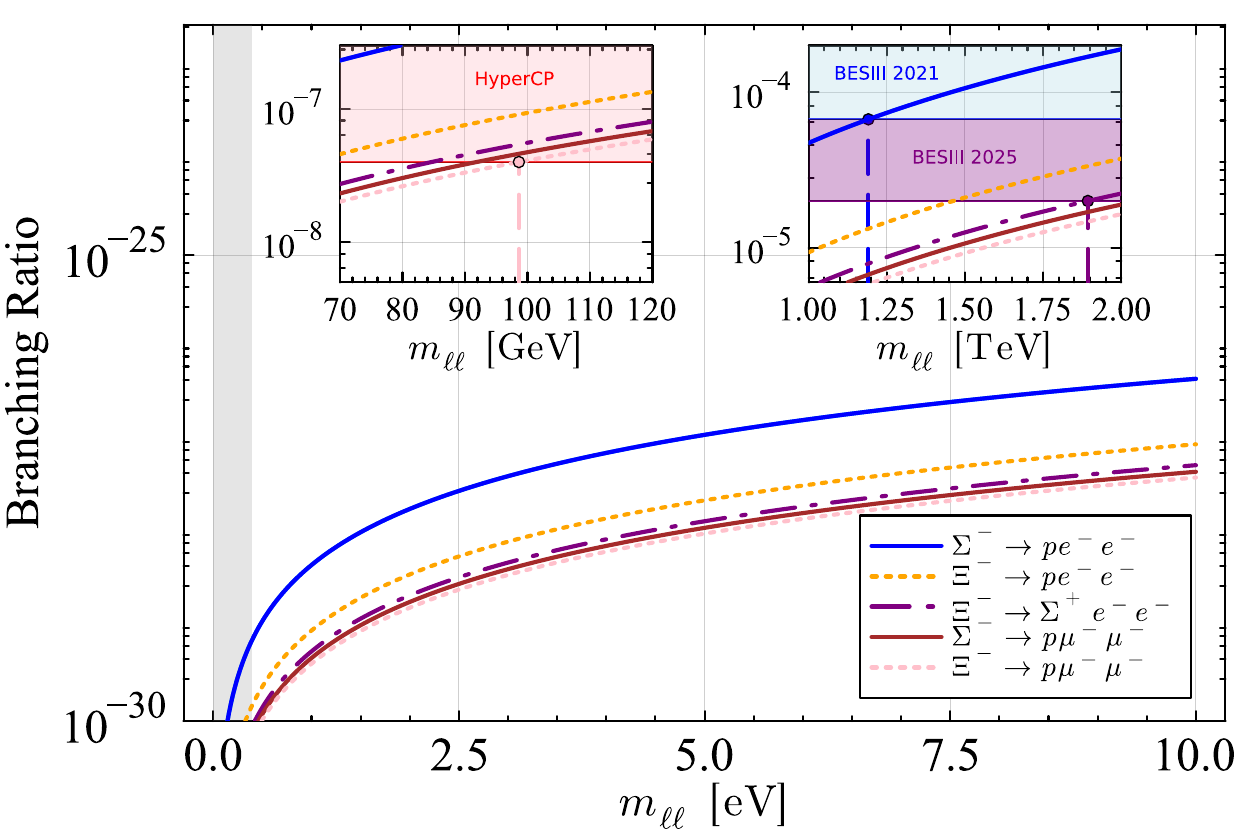}
\caption{Dependence of the LNV hyperon decay branching ratios, obtained from the one-loop amplitudes in the EOMS scheme with the renormalization scale $\mu=m$, on the effective Majorana mass $m_{\ell\ell}$. The insets compare the upper limits from HyperCP (left) \cite{HyperCP:2005sby} and BESIII (right) \cite{BESIII:2020iwk,BESIII:2025ylz}. A zoomed-in view of the gray band, which covers $0 \leq m_{\ell\ell} \leq 400$~meV, is provided in figure~\ref{fig:Br}.}
\label{fig:Br1}
\end{figure}

It is also interesting to analyze the dependence of the branching ratios for the LNV hyperon decays on the effective Majorana neutrino mass, shown in figure~\ref{fig:Br1}. The branching ratios increase only mildly with varying $m_{\ell\ell}$ and remain below the order of $10^{-25}$ for $m_{\ell\ell} < 10~{\rm eV}$, posing a significant challenge for direct searches for light Majorana neutrinos at colliders. As illustrated by the insets, to reach the HyperCP upper limit~\cite{HyperCP:2005sby}, 
one would need $m_{\mu\mu}\sim 100\ \mathrm{GeV}$ if the experimental limit is naively translated within the light-neutrino-exchange parametrization. Similarly, saturating the BESIII bounds~\cite{BESIII:2020iwk,BESIII:2025ylz} would correspond to $m_{ee}$ at the level of around 1 TeV within the same naive mapping. However, since our BChPT model for $0\nu\beta\beta$ decays is valid only for light-neutrino exchange, these translated scales should not be interpreted as evidence for heavy neutrinos, but rather as indicating that a signal near current bounds would require mechanisms beyond light-neutrino exchange (e.g., heavy-neutrino exchange or short-range operators). For the incorporation of heavy neutrinos in the exchange mechanism, we refer the reader to, e.g., refs.~\cite{Zhou:2021lnl,Chen:2025svf}.

\begin{figure}[tb]
\centering
\includegraphics[width=0.7\textwidth]{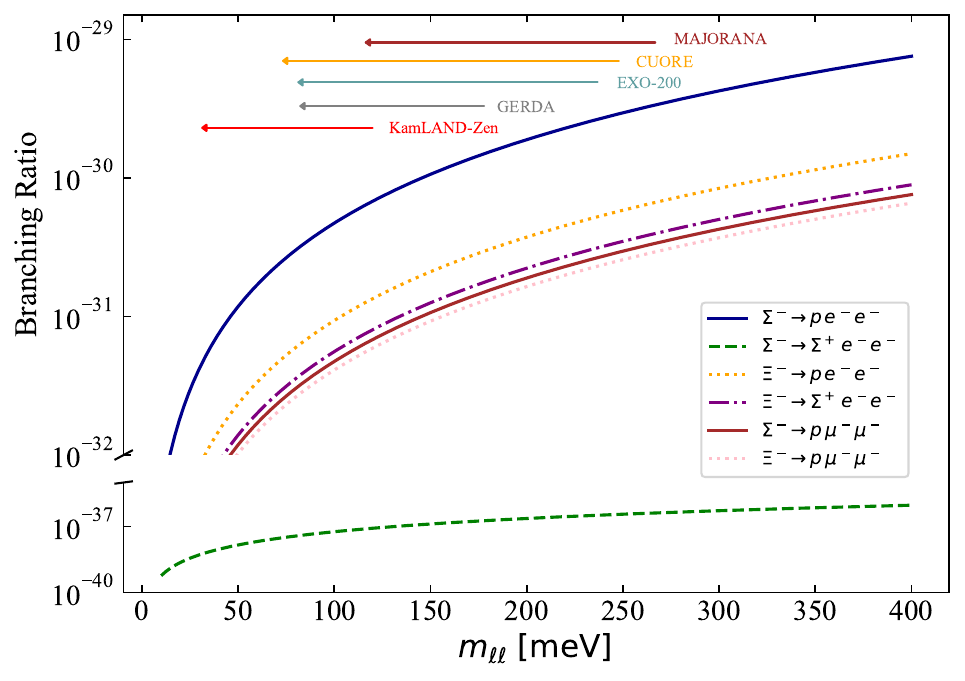}
\caption{Branching ratios of hyperon $0\nu\beta\beta$ decays, obtained from the one-loop amplitudes in the EOMS scheme with the renormalization scale $\mu=m$, as functions of the effective Majorana mass $m_{\ell\ell}$. The curves span the range $0\leq m_{\ell\ell}\leq 400$~meV, covering most of the recent experimental constraints: KamLand-Zen~\cite{KamLAND-Zen:2024eml}, CUORE~\cite{CUORE:2024ikf}, \textsc{Majorana}~\cite{Majorana:2022udl}, GERDA~\cite{GERDA:2020xhi} and EXO-200~\cite{EXO-200:2019rkq} .}
\label{fig:Br}
\end{figure}

To date, various nuclear $0\nu\beta\beta$ measurements have led to stringent limit on the effective Majorana mass of the electronic neutrino. In particular, the KamLAND-Zen has reached the inverted mass ordering region where $m_{ee}<50$~meV is demanded~\cite{KamLAND-Zen:2022tow}. Therefore, we zoom in the meV region of $m_{\ell\ell}$, indicated by the gray band in figure~\ref{fig:Br1} and enlarged in figure~\ref{fig:Br}. In this figure, our BChPT predictions are shown for $m_{\ell\ell}\leq 400$~meV, together with the experimental constraints of $m_{ee}$ from~\cite{KamLAND-Zen:2024eml,CUORE:2024ikf,Majorana:2022udl,GERDA:2020xhi,EXO-200:2019rkq}.
The theoretical branching ratio for the $\Sigma^- \to \Sigma^+ e^- e^-$ decay stays below $10^{-36}$ throughout the range of $m_{\ell\ell}\leq 400$~meV, while for the other processes the magnitudes of the branching ratio are typically of the order $10^{-29}$. Therefore, under the assumption that the decay is dominated by light Majorana neutrino exchange mechanism, the hyperon $0\nu\beta\beta$ decays are not expected to be observable at current and future colliders. 

To obtain quantitative predictions at the $\mathcal{O}(p^2)$ order, it is necessary to know the values of the unknown LECs $g_i$ and $g_i^\prime$ (c.f.~\eqref{eq.counterterm}). In the next subsection, we are going to discuss the possibility of accessing the hyperon $0\nu\beta\beta$ decays through lattice QCD simulation of neutrinoless TFFs.

\subsection{Neutrinoless transition form factor}
\label{sec:TFF}

The leptonic tensor in eq.~\eqref{eq:Lmunu} can be further decomposed into 
\begin{align}
    &L^{\mu\nu} = L^{\mu\nu}_S+ L^{\mu\nu}_A\ ,\\
   & L^{\mu\nu}_S\equiv \bar{u}_{\ell\,L}(k_1)g^{\mu\nu} C \bar{u}_{\ell\,L}^T(k_2)\ ,\quad
    L^{\mu\nu}_A\equiv -{\rm i}\bar{u}_{\ell\,L}(k_1)\sigma^{\mu\nu} C \bar{u}_{\ell\,L}^T(k_2)\ .
\end{align}
with $\sigma^{\mu\nu}=\frac{\rm i}{2}[\gamma^{\mu},\gamma^{\nu}]$. In the case where the four momenta of the final leptons are identical $k_1=k_2\equiv k$, the hadronic tensor~\eqref{eq.hadron.tensor.total} is symmetric under the permutation of the Lorentz indices $\mu$ and $\nu$, while $L^{\mu\nu}_A$ is antisymmetric in $\mu$ and $\nu$. Hence, the amplitude proportional to $L^{\mu\nu}_A$ vanishes, leading to 
\begin{align}
     \mathcal{M}_{0\nu\beta\beta}&=  C_{\rm Lept}\, {H}_{\mu\nu}\, L^{\mu\nu}_S \notag\\
     &= C_{\rm Lept} \big\{\bar{u}(p_2)\left[\mathcal{S}(s)+\gamma_5 \mathcal{P}(s)\right]u(p_1) \big\}\big\{\bar{u}_{\ell L}(k) C \bar{u}_{\ell L}^T(k)\big\}\ .
\end{align}
In deriving the above expression, we have used the Gordon identities to reduce the baryonic bilinears to the scalar and pseudoscalar structures $\bar{u}(p_2)u(p_1)$ and $\bar{u}(p_2)\gamma_5 u(p_1)$.
We define scalar functions $\mathcal{S}(s)$ and $\mathcal{P}(s)$ as neutrinoless TFFs, arising from the matrix element of an effective neutrinoless weak current of momentum $2k$ sandwiched between two baryon states. These neutrinoless TFFs encode information about the underlying strong dynamics and, in principle, can be accessed by lattice QCD in the future. In the degenerate baryon mass limit, the presence of $\gamma_5$ between the two baryon spinors results in an identically zero contribution. Therefore, in the following, we will focus only on the $\mathcal{S}(s)$ form factor.

The TFF $\mathcal{S}(s)$ can be expressed in terms of the original \(V_i\) and \(A_i\) by the following relation
\begin{align}
\mathcal{S}(s) =&\, d(V_1+V_4)+\frac{s-2m_{\ell}^2}{2}(V_5+V_6)+\frac{m_{\ell}^2}{2}(2V_7+2V_8+V_{10}-V_{11}+V_{12}-V_{13}) +m^2_{1}V_9\notag\\
&+\frac{m_{1}^2-u}{2}(V_{10}+V_{12}+2V_{29}+2V_{30}+2V_{32}+2V_{33}) +\frac{s+u-m_{2}^2}{2}(V_{11}+V_{13})\notag\\
&+m_{1}(V_{23}+V_{24}) +(m_{1}-m_{2})(V_{26}+V_{28})+m_{\ell}^2(V_{29}+V_{30}+V_{31}+V_{34}) \ ,
\end{align}
where $t=u=(\Sigma_m-s)/2$ always holds in the case of $k_1=k_2$. For degenerate baryon masses, the explicit expression of $\mathcal{S}(s)$ can be written as
\begin{align}
\label{eq:TFFS}
\mathcal{S}(s) = 8 \tilde{g}_1 + \widetilde{ \mathcal{S}}(s)^{\rm loop}\ ,
\end{align}
where the first term originates from the local counterterm Lagrangian and the second term represents the one-loop contribution. Explicit expressions for the one-loop TFFs $\widetilde{\mathcal{S}}(s)^{\rm loop}$ are provided in appendix~\ref{app:TFF_expressions}. The overtilde indicates that the chiral result $\mathcal{S}(s)$ has been EOMS-renormalized. The LEC $\tilde{g}_1$ is an unknown constant that parameterizes physics at distance scales on the order of or shorter than the inverse of the hard momentum scale in BChPT. 
As a Wilson coefficient, its value may be estimated by the naturalness ansatz in effective field theories. 
A more practical way of deterimining the LECs is to employ the state-of-the-art lattice QCD calculations. In lattice QCD, the neutrinoless mesonic process $\pi^-\to \pi^+ e^-e^-$ (or $\pi^- \pi^-\to  e^-e^-$ by crossing) has been extensively investigated, e.g., in Refs.~\cite{Nicholson:2018mwc,Feng:2018pdq,Tuo:2019bue,Detmold:2020jqv,Detmold:2022jwu,Boyle:2025vwt}. The hyperon $0\nu\beta\beta$ decays, particularly the TFF $\mathcal{S}(s)$, can also be simulated on the lattice, such that the LEC $\tilde{g}_1$ can be determined to evaluate the size of the LO short-range contribution. Below, we explore the $s$-dependence and pion mass dependence of the neutrinoless TFFs $\mathcal{S}(s)$ based on their chiral expressions.

\begin{figure}[tb]
\centering
\includegraphics[width=0.998\linewidth]{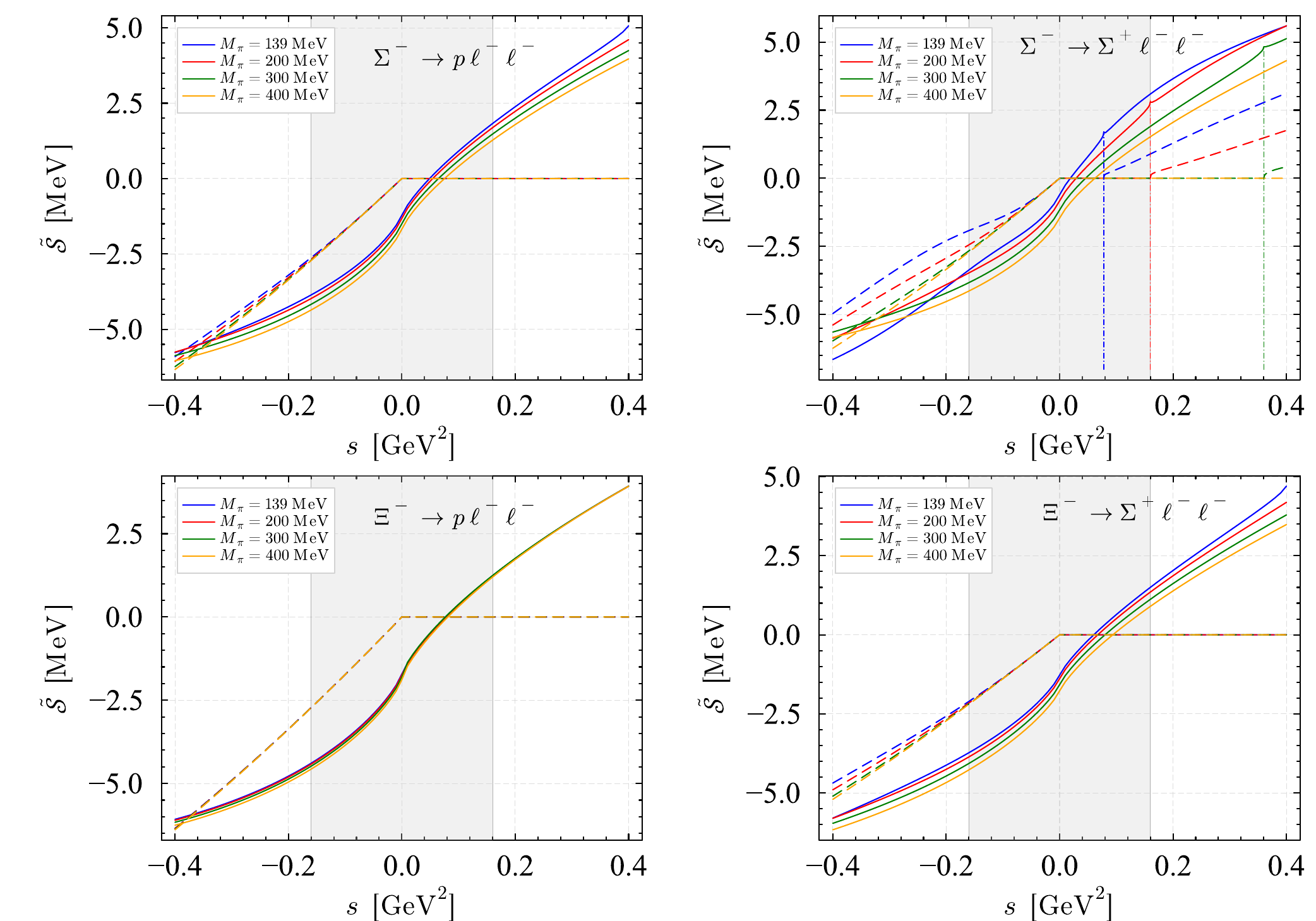}
\caption{Neutrinoless TFFs as functions of the Mandelstam variable \(s\), calculated with four different pion masses: $M_{\pi}=139$, $200$, $300$ and $400$~MeV. Real and imaginary parts are shown as solid and dashed lines, respectively. In the top-right panel, the vertical lines indicate the two-Goldstone-boson thresholds. The final leptons are electrons with zero mass. The gray shaded region indicates the nominal validity domain of SU(3) BChPT ($-0.16\leq s \leq 0.16~{\rm GeV}^2$).}
\label{fig:sruning}
\end{figure}

Figure~\ref{fig:sruning} presents the $s$-dependence of the TFF $\widetilde{\mathcal{S}}(s)^{\rm loop}$ for the four physical processes in the domain $-0.4\leq s \leq 0.4$~GeV$^2$. In our numerical analysis, all the baryon masses are taken to be the chiral limit mass $m=1$~GeV and the lepton masses are neglected, i.e., $m_{\ell}=0$.\footnote{In fact, the analytical expressions of $\Sigma^- \to p\, \ell^- \ell^-$ and $\Xi^- \to \Sigma^+ \ell^- \ell^-$ are identical, as can be seen from eq.~\eqref{eq:S_Spll}. 
In order to assess the influence of baryon mass, we set $m=1$~GeV for the former and $m=(m_{\Xi^-}+m_{\Sigma^-})/2$ for the latter.} 
The BChPT calculation is expected to be valid only in the low-momentum region; accordingly, the domain $-0.16\leq s \leq 0.16$~GeV$^2$ is indicated by the gray shaded region. 
It can be found that the TFFs always possess imaginary parts in the negative-$s$ regime. The occurrence of these imaginary parts is due to the fact that the internal baryon and the neutrino (c.f. diagrams (a), (c), (d) and (f) in figure~\ref{fig.Feynman.Diagram}) can be simultaneously put on-shell when $u=t=m^2-s/2>m^2$ for $s<0$. 
In the timelike region $s\geq0$, the onset of the TFF imaginary part occurs at the threshold of the two intermediate Goldstone bosons appearing in diagram (d) of figure~\ref{fig.Feynman.Diagram}. This feature is most clearly seen in the $\Sigma^-\to \Sigma^+\ell\ell$ process, where the imaginary part appears once $s>s_{\rm th}=(2M_{\pi})^2$.  For the other decays, the thresholds are located beyond $0.4$~GeV$^2$. Our results suggest that it is preferable to conduct lattice QCD simulations in the region $0\leq s\leq s_{\rm th}$ where the neutrinoless TFFs are real.

We proceed to investigate the behavior of the TFF $\widetilde{\mathcal{S}}(s)^{\rm loop}$ by varying the $u/d$-quark masses, $m_{u/d}$, while keeping the strange quark mass $m_s$ fixed. In practice, the quark masses can be associated with the LO Goldstone boson masses in ChPT by~\cite{Gasser:1984gg}
\begin{align}\label{eq.GB.masses.LO}
M_{\pi}^2 =2B_0\hat{m}\ ,\quad M_{K}^2 =B_0(\hat{m}+m_s)\ ,\quad
 M_{\eta}^2 ={2}B_0(\hat{m}+2m_s)/3\ ,
\end{align}
satisfying the well-known Gell-Mann-Okubo relation $4M_K^2=3M_{\eta}^2+M_{\pi}^2$. Here $\hat{m}=(m_u+m_d)/2$ is the average of the light $u/d$-quark masses. The pion-mass (or $u/d$-quark mass) dependence of $M_K$ can be obtained from eq.~\eqref{eq.GB.masses.LO}, which reads $M_K^2={M_\pi^2}/{2}+\overline{M}_K^2$ with $\overline{M}_K^2=B_0m_s$ standing for the kaon mass in the SU(2) chiral limit, i.e., $(m_u,m_d)\to 0$. We determine the value of $\overline{M}_K^2$ by using the physical masses of the pion and kaon as given in table~\ref{tab:octet_masses}. The pion mass dependence of the baryon masses is ignored. In figure~\ref{fig:sruning}, we compare the TFFs obtained with four different pion masses: $M_{\pi}=139, 200, 300, 400$~MeV. It can be found that, at a fixed $s$, the effect of varying the pion mass in the TFFs is marginal.

\section{Summary and Outlook}
\label{sec:summary}

We present a systematic study of $0\nu\beta\beta$ decays of spin-$1/2$ hyperons. Our analysis is performed within the covariant framework of SU(3) BChPT and focuses on the long-range mechanism mediated by light Majorana neutrino exchange. The decay amplitudes, arising from the low-energy realization of the dimension-five Weinberg LNV operator, first appear at one-loop level, i.e., $\mathcal{O}(p^3)$. The loop amplitudes are renormalized using DR with the $\overline{\rm MS}-1$ subtraction scheme together with the EOMS scheme, which removes both UV divergences and PCB terms. The resulting chiral expressions satisfy the correct power counting and preserve analyticity, thereby ensuring reliable predictions with controlled theoretical uncertainties.

Using the EOMS renormalized one-loop amplitudes, we have computed the differential decay rates and branching ratios for all kinematically allowed hyperon $0\nu\beta\beta$ decay channels. 
These branching ratios correspond to the long-range one-loop contribution from light-neutrino exchange; the overall rates also depend on the short-range counterterm LECs, which remain unknown at present.
Our numerical results show that the branching ratios are tiny as expected, more than 20 orders of magnitude smaller than the current experimental upper bounds.

The results mean that any possible nonvanishing branching fractions that may be observed would indicate new physics beyond the light-neutrino-exchange mechanism.
Furthermore, local counterterms responsible for the short-distance $\Delta L=2$ contribution start at $\mathcal{O}(p^2)$. We propose that this contribution can be determined through lattice QCD calculations of the neutrinoless transition form factors. The present study provides model-independent results that will be useful for future experimental searches for LNV signatures via hyperon $0\nu\beta\beta$ decays, complementing nuclear fixed-target experiments.

\acknowledgments
We would like to thank Prof. Hai-Bo Li for reading the manuscript and for providing valuable comments. DLY appreciates the hospitality of the Institute of Theoretical Physics (ITP) at Chinese Academy of Sciences (CAS), where part of this work was done. This work is supported by National Nature Science Foundations of China under Grants No.~12547166, No.~12275076, No.~12335002, No.~12125507, No.~12447101, and No.~12505105; by the Science Fund for Distinguished Young Scholars of Hunan Province under Grant No.~2024JJ2007; by the Fundamental Research Funds for the Central Universities under Grant No.~531118010379; by the Science Foundation of Hebei Normal University under Grants No. L2025B09; by Science Research Project of Hebei Education Department under Grant No. QN2025063; by Hebei Natural Science Foundation under Grant No. A2025205018; and by CAS under Grant No. YSBR-101.


\appendix

\section{Reduction of the leptonic part}
\label{app:lept}

In the energy-momentum representation, the leptonic part can be derived as
\begin{align}
\mathcal{L}^{\mu\nu}(k_1,k_2;q) &=\sum_{s,s^
\prime}\big[-2\sqrt{2}G_F \bar{u}_{\ell\,L}(k_1)\gamma^{\mu} \upsilon^{(s)}_{\ell\,L}(q)\big]\frac{\rm i}{q^2+{\rm i}\epsilon}\big[\bar{\upsilon}^{(s)}_{\ell\,L}(q)(-\frac{\rm i}{2}m_{\ell\ell}C)\bar{\upsilon}^{(s^{\prime}),T}_{\ell\,L}(-q)\big]\notag\\
&\times\frac{\rm i}{(-q)^2+{\rm i}\epsilon}\big[-2\sqrt{2}G_F \bar{u}_{\ell\,L}(k_2)\gamma^{\nu} \upsilon^{(s^{\prime})}_{\ell\,L}(-q)\big]\ .\label{eq.leptonic.tensor.original}
\end{align}
Since we are interested in the light-neutrino exchange, the mass in the neutrino propagator is omitted, following ref.~\cite{Cirigliano:2017tvr}. The summation is taken over the neutrino spins $s$ and $s^\prime$. The term in the last bracket is just a $c$-number, which remains the same by taking the transpose. Therefore, the above equation is recast into the form as
\begin{align}
\mathcal{L}^{\mu\nu}(k_1,k_2;q) &=(-2\sqrt{2}G_F)^2  \bar{u}_{\ell\,L}(k_1)\gamma^{\mu}\frac{{\rm i}\sum_s \upsilon_{\ell\,L}^{(s)}(q)  \bar{\upsilon}_{\ell\,L}^{(s)} (q) }{q^2+{\rm i}\epsilon}(-\frac{\rm i}{2}m_{\ell\ell}C)\notag\\
 &\times\frac{{\rm i}\sum_{s^\prime}\bar{\upsilon}^{(s^\prime),T}_{\ell\,L}(-q)\upsilon^{(s^\prime),T}_{\ell\,L}(-q)}{(-q)^2+ {\rm i}\epsilon}\gamma^{\nu\,T} \bar{u}_{\ell\,L}^{T}(k_2)\ .
\end{align}
For the above equation, one can apply the formulae of polarization sums
\begin{align}
  \sum_s \upsilon_{\ell\,L}^{(s)}(q)  \bar{\upsilon}_{\ell\,L}^{(s)} (q) = \slashed{q}\ ,\quad \sum_{s^\prime }\bar{\upsilon}^{(s^\prime),T}_{\ell\,L}(-q)\upsilon^{(s^\prime),T}_{\ell\,L}(-q)=-\slashed{q}^T\ ,
\end{align}
where $\slashed{q}^T=-C\slashed{q}C^{-1}$ and $C={\rm i}\gamma^2\gamma^0$ is the charge conjugation matrix. 
One finds $\slashed{q}(-\slashed{q}^T)=q^2$, which cancels one power of the propagator and leaves a single scalar propagator $S(q^2)$.
Then, the leptonic tensor is simplified to
\begin{align}\label{eq.LL.eq}
   \mathcal{L}^{\mu\nu}(k_1,k_2;q) &=\bigg[\frac12 m_{\ell\ell}(-2\sqrt{2}G_F)^2V_{ui}V_{uj}\bigg]\frac{\rm i}{q^2+{\rm i}\epsilon}
   \bigg[\bar{u}_{\ell\,L}(k_1)\gamma^{\mu}\gamma^{\nu} C \bar{u}_{\ell\,L}^T(k_2)\bigg]\ .
\end{align}   
We denote the coefficient in first bracket by $C_{\rm Lept}$, the propagator by $S(q^2)$, and the term in the last bracket by $L^{\mu\nu}$. Equation~\eqref{eq.LL.eq} turns to
\begin{align}
\mathcal{L}^{\mu\nu}(k_1,k_2;q)=C_{\rm Lept}\, S(q^2)\,  {L}^{\mu\nu}(k_1,k_2)\ .
\end{align}
with
\begin{align}
    C_{\rm Lept} &=\frac{1}{2}m_{\ell\ell}(-2\sqrt{2}G_F)^2V_{ui}V_{uj}\ ,\\
    S(q^2)& = \frac{\rm i}{q^2+{\rm i}\epsilon}\ ,\\
   {L}^{\mu\nu}(k_1,k_2)&= \bar{u}_{\ell\,L}(k_1)\gamma^{\mu}\gamma^{\nu} C \bar{u}_{\ell\,L}^T(k_2)\ . 
\end{align}
For clarity, the rank-2 Lorentz structure $L^{\mu\nu}$ is usually called leptonic tensor. The leptonic tensor possesses the following property
\begin{align}\label{eq:su}
    {L}^{\nu\mu}(k_2,k_1)=-{L}^{\mu\nu}(k_1,k_2) \ , 
\end{align}
or, more transparently,
\begin{align}
   \bar{u}_{\ell\,L}(k_2)\gamma^{\nu}\gamma^{\mu} C\bar{u}_{\ell\,L}^T(k_1)=-\bar{u}_{\ell\,L}(k_1)\gamma^{\mu}\gamma^{\nu} C \bar{u}_{\ell\,L}^T(k_2)\ ,
\end{align}
which is useful in deriving the $t$-channel hadronic amplitudes from the $u$-channel ones.

\section{Lorentz operators for chiral expansion}
\label{app:operators}
The Lorentz tensor operators defined in eq.~\eqref{tab:operators} and eq.~\eqref{tab:operators_OA} are complete but not independent. Here, we only discuss the redundancy caused by the contraction of the leptonic tensor $L^{\mu\nu}$~\eqref{eq:Lmunu}. Taking the $\mathcal{O}_{V,\mu\nu}^6=k_{1\nu}k_{2\mu}$ operator for example, it is straightforward to obtain the following identity
\begin{align}
    \big[\bar{u}(p_2)\mathcal{O}_{V,\mu\nu}^6 u(p_1)\big]\,L^{\mu\nu}
    =\bigg[\bar{u}(p_2)\big(\frac{s-2m_{\ell}^2}{d}\mathcal{O}_{V,\mu\nu}^1-\mathcal{O}_{V,\mu\nu}^5 \big)  u(p_1) \bigg]\,L^{\mu\nu}\ ,
\end{align}
where $\mathcal{O}_{V,\mu\nu}^1=g_{\mu\nu}$ and $\mathcal{O}_{V,\mu\nu}^5=k_{1\mu}k_{2\nu}$.
This indicates that $\mathcal{O}_V^6$ can be expressed in terms of $\mathcal{O}_V^1$ and $\mathcal{O}_V^5$. Namely, the $\mathcal{O}_V^6$ operator is redundant and can be eliminated by $\mathcal{O}_{V}^6  \rightarrow ({s-2 m_{\ell}^2})\mathcal{O}_{V}^1/{d} - \mathcal{O}_{V}^5$. In the same manner, we have obtained all the possible replacement rules, which are summarized below:
\begin{align}
\mathcal{O}_{V,A}^6 & \rightarrow \frac{s-2 m_{\ell}^2}{d}\mathcal{O}_{V,A}^1 - \mathcal{O}_{V,A}^5\ , \\
\mathcal{O}_{V,A}^{7,8} & \rightarrow \frac{m_{\ell}^2}{d}\mathcal{O}_{V,A}^1\ , \qquad
\mathcal{O}_{V,A}^{9}  \rightarrow \frac{m^2}{d}\mathcal{O}_{V,A}^1\ , \\
\mathcal{O}_{V,A}^{12} & \rightarrow \frac{m_{\ell}^2+m^2-u}{d}\mathcal{O}_{V,A}^1 - \mathcal{O}_{V,A}^{10}\ , \\
\mathcal{O}_{V,A}^{13} & \rightarrow \frac{m_{\ell}^2+m^2-t}{d}\mathcal{O}_{V,A}^1 - \mathcal{O}_{V,A}^{11}\ , \\
\mathcal{O}_{V,A}^{15} & \rightarrow \frac{s-2 m_{\ell}^2}{d}\mathcal{O}_{V,A}^2 - \mathcal{O}_{V,A}^{14}\ , \\
\mathcal{O}_{V,A}^{16,17} & \rightarrow \frac{m_{\ell}^2}{d}\mathcal{O}_{V,A}^2\ , \qquad
\mathcal{O}_{V,A}^{18}  \rightarrow \frac{m^2}{d}\mathcal{O}_{V,A}^2\ , \\
\mathcal{O}_{V,A}^{21} & \rightarrow \frac{m_{\ell}^2+m^2-u}{d}\mathcal{O}_{V,A}^2 - \mathcal{O}_{V,A}^{19}\ , \\
\mathcal{O}_{V,A}^{22} & \rightarrow \frac{m_{\ell}^2+m^2-t}{d}\mathcal{O}_{V,A}^2 - \mathcal{O}_{V,A}^{20}\ , \\
\mathcal{O}_{V}^{24} & \rightarrow \frac{2 m}{d}\mathcal{O}_{V}^1 - \mathcal{O}_{V}^{23}\ , \qquad
\mathcal{O}_{A}^{24} \rightarrow -\frac{2 m}{d}\mathcal{O}_{A}^1 - \mathcal{O}_{A}^{23}\ , \\
\mathcal{O}_{V,A}^{27} & \rightarrow \frac{2}{d}\mathcal{O}_{V,A}^2 - \mathcal{O}_{V,A}^{25}\ , \\
\mathcal{O}_{V}^{28} & \rightarrow -\frac{2}{d}\mathcal{O}_{V}^2 - \mathcal{O}_{V}^{26}\ , \qquad
\mathcal{O}_{A}^{28} \rightarrow -\frac{4m}{d}\mathcal{O}_{A}^1 - \frac{2}{d}\mathcal{O}_A^2 - \mathcal{O}_{A}^{26}\ , \\
\mathcal{O}_{V}^{30} & \rightarrow \frac{2(m_{\ell}^2+m^2-u)}{d}\mathcal{O}_{V}^1 - \frac{2 m}{d}\mathcal{O}_{V}^2 - \mathcal{O}_{V}^{29}\ , \\
\mathcal{O}_{A}^{30} & \rightarrow \frac{2(m_{\ell}^2+m^2-u)}{d}\mathcal{O}_{A}^1 + \frac{2 m}{d}\mathcal{O}_{A}^2 - \mathcal{O}_{A}^{29}\ , \\
\mathcal{O}_{V}^{32} & \rightarrow \frac{2(m^2-u)}{d}\mathcal{O}_{V}^1 - \frac{4 m}{d}\mathcal{O}_{V}^2 - \mathcal{O}_{V}^{33}\ , \\
\mathcal{O}_{A}^{32} & \rightarrow \frac{2(m^2-u)}{d}\mathcal{O}_{A}^1 - \mathcal{O}_{A}^{33}\ , \\
\mathcal{O}_{V,A}^{34} & \rightarrow \frac{2 m_{\ell}^2}{d}\mathcal{O}_{V,A}^1 - \mathcal{O}_{V,A}^{31}\ .
\end{align}
In the above, the Lorentz indices of the operators are suppressed for simplicity. The set of Lorentz operators is reduced to 
\begin{align}
\{\mathcal{O}_{V,A}^{i}\ , i=1-5,10,11,14,19,20,23,25,26,29,31,33\} \ .
\end{align}
It is worth noting that the reduced set of Lorentz operators is suitable for performing chiral expansion. Therefore, we have applied them in extracting the PCB terms in the procedure of renormalization, as shown section~\ref{sec:Renormalization}.

\section{One-loop expressions of hadronic tensors}
\label{app:hadronic_Amp}

In the SU(3) Cartesian basis, the $0\nu\beta\beta$ decays can be uniformly written in the form as
\begin{align}
B_i^-(p_1) \to B_j^+(p_2)\,\ell_a^-(k_1)\,\ell_b^-(k_2)\ ,
\end{align}
where $i,j,a,b\in\{ 1,\ldots,8\}$. The indices $a,b$ of the final leptons inherit those of the charged weak currents by virtue of eq.~\eqref{eq.landr} and eq.~\eqref{eq:T}. The use of a Cartesian basis results in a universal SU(3) decay amplitude (c.f. eq.~\eqref{eq.SU3.amplitude}), from which the physical amplitudes for any given real processes can be deduced (cf. eq.~\eqref{eq:Amp_phys_Xpll}). 

For the diagrams in figure~\ref{fig.Feynman.Diagram}, the relevant one-loop amplitudes obtained in the Cartesian basis are listed below.
\begin{itemize}
\item Diagram (a)
\begin{align}
{H}^{\mu\nu}_{a[ij,ab]} &= \frac{{\rm i}}{8}f_{lja}({\rm i}D\, d_{ilb}-F\,f_{ilb})\times \int\frac{{\rm d}^d k}{(2\pi)^d}\frac{\gamma^\nu (k^2+m_l \slashed{k}+\slashed{p}_1 \slashed{k})\gamma_5 k^{\mu}}{[k^2-M_b^2][(k+p_1)^2-m_l^2][(k+k_1)^2]}\notag\\
&+(-\frac{{\rm i}}{8})({\rm i}D^2\,d_{ilb}d_{lja}-D\,F\,d_{ilb}f_{lja}-D\,F\,f_{ilb}d_{lja}-{\rm i}F^2\,f_{ilb}f_{lja})\notag\\
&\times\int\frac{{\rm d}^d k}{(2\pi)^d}\frac{\gamma^\nu (k^2-m_l\slashed{k}+\slashed{p}_1\slashed{k})k^{\mu} }{[k^2-M_b^2][(k+p_1)^2-m_{l}^2][(k+k_1)^2]}\ .
\end{align}
\item Diagram (b)
\begin{align}
{H}^{\mu\nu}_{b[ij,ab]} &=(-\frac{\sqrt{2}}{8})f_{abe}f_{ije}\times  \int \frac{{\rm d}^d k}{(2\pi)^d} \frac{\gamma^{\nu}k^{\mu}}{[k^2-m_a^2][(k+k_1)^2]}\notag\\
&+\frac{\sqrt{2}}{8}f_{abe}(D\, d_{ije}-F\, f_{ije})\times \int \frac{{\rm d}^d k}{(2\pi)^d} \frac{\gamma^{\nu}k^{\mu}\gamma_5 }{[k^2-m_a^2][(k+k_1)^2]}\ .
\end{align}
\item Diagram (c)
\begin{align}
{H}^{\mu\nu}_{c[ij,ab]} &=\frac{{\rm i}}{8}(D\, d_{lja}+{\rm i}F\, f_{lja})f_{ilb}\times  \int \frac{{\rm d}^d k}{(2\pi)^d} \frac{\slashed{k} \gamma_5(\slashed{p}_2-\slashed{k}+m_l)\gamma^{\mu}k^{\nu}}{[k^2-m_a^2][(p_2-k)^2-m_l^2][(k+k_2)^2]}\notag\\
&-\frac{{\rm i}}{8}({\rm i}D^2\, d_{lja}d_{ilb}-D\, F\, d_{lja}f_{ilb}-D\, F\, f_{lja}d_{ilb}-{\rm i}F^2\, f_{lja}f_{ilb}) \notag\\
&\times\int \frac{{\rm d}^dk}{(2\pi)^d} \frac{\slashed{k} (\slashed{p}_2-\slashed{k}-m_l)\gamma^{\mu}k^{\nu}}{[k^2-m_a^2][(p_2-k)^2-m_l^2][(k+k_2)^2]}\ .
\end{align}
\item 
Diagram (d)
\begin{align}
{H}^{\mu\nu}_{d[ij,ab]} &=\frac{1}{16}(D^2\, d_{ljb}d_{ila}+{\rm i}D\, F\, d_{ljb}f_{ila}+{\rm i}D\, F\, f_{ljb}d_{ila}-F^2\, f_{ljb}f_{ila}) \\
&\times  \int \frac{{\rm d}^d k}{(2\pi)^d} \frac{(\slashed{p}_2-\slashed{k}-\slashed{p}_1)(\slashed{k}+\slashed{p}_1-m_l)\slashed{k} k^{\mu} (k+p_1-p_2)^{\nu}}{[(k+p_1)^2-m^2_l][k^2-m_a^2][(p_2-k-p_1)^2-m_b^2][(k+k_1-p_1)^2]}\ .\notag
\end{align}
\item Diagram $(e)$:
\begin{align}
{H}^{\mu\nu}_{e[ij,ab]} &=\frac{{\rm i}}{32}f_{ije}f_{abe}\times  \int \frac{{\rm d}^d k}{(2\pi)^d} \frac{(2\slashed{k}-\slashed{p}_2+\slashed{p}_1)k^{\mu} (k+p_1-p_2)^{\nu}}{[k^2-m_a^2][(p_2-k-p_1)^2-m_b^2][(k+k_1)^2]}\ .
\end{align}
\item
Diagram $(f)$:
\begin{align}
{H}^{\mu\nu}_{f[ij,ab]} &= (-\frac{{\rm i}}{4})f_{lja}f_{ilb}\times \int \frac{{\rm d}^dk}{(2\pi)^d}\frac{\gamma^{\nu}[(\slashed{k}-\slashed{k}_1+\slashed{p}_1)+m_l]\gamma^{\mu}}{[(k-k_1+p_1)^2-m_l^2][k^2]}\notag\\
&+\frac{1}{4}f_{lja}(D\,d_{ilb}+{\rm i}F\,f_{ilb})\times \int \frac{{\rm d}^dk}{(2\pi)^d}\frac{\gamma^{\nu}[(\slashed{k}-\slashed{k}_1+\slashed{p}_1)+m_l]\gamma^{\mu}\gamma_5}{[(k-k_1+p_1)^2-m_l^2][k^2]}\notag\\
&+\frac{1}{4}f_{ilb}(D\,d_{lja}+{\rm i}F\,f_{lja})\times \int \frac{{\rm d}^dk}{(2\pi)^d}\frac{\gamma^{\nu}[(\slashed{k}-\slashed{k}_1+\slashed{p}_1)-m_l]\gamma^{\mu}\gamma_5}{[(k-k_1+p_1)^2-m_l^2][k^2]}\notag\\
&-\frac{1}{4}({\rm i}D^2\, d_{lja}d_{ilb}-D\, F\, d_{lja}f_{ilb}-D\, F\, f_{lja}d_{ilb}-{\rm i}F^2\, f_{lja}f_{ilb})\notag\\
&\times \int \frac{{\rm d}^dk}{(2\pi)^d}\frac{\gamma^{\nu}[(\slashed{k}-\slashed{k}_1+\slashed{p}_1)-m_l]\gamma^{\mu}}{[(k-k_1+p_1)^2-m_l^2][k^2]}\ .
\end{align}
\end{itemize}
where summation over the internal index $l=1,2,\cdots,8$ is implied,  and $\lambda_a(a=1,\ldots,8)$ are the Gell-Mann matrices.
The $f_{abc}$ and $d_{abc}$ are structure constants of the $su(3)$ algebra. Their values can be computed from the Gell-Mann matrices via
$f_{abc}=\frac{1}{4i}\mathrm{Tr}([\lambda_a,\lambda_b]\lambda_c)$ and $d_{abc}=\frac{1}{4}\mathrm{Tr}(\{\lambda_a,\lambda_b\}\lambda_c)$. Furthermore, $M_{1-3}=M_\pi$, $M_{4-7}=M_K$ and $M_{8}=M_\eta$ for the masses involved in the internal propagators of the Goldstone bosons. The internal baryon mass $m_l$ can be taken as the mass $m$ in the SU(3) chiral limit. 

The total hadronic amplitude is 
\begin{align}\label{eq.SU3.amplitude}
    {H}^{\mu\nu}_{ij,ab} = {H}^{\mu\nu}_{a[ij,ab]} + {H}^{\mu\nu}_{b[ij,ab]}+{H}^{\mu\nu}_{c[ij,ab]}+{H}^{\mu\nu}_{d[ij,ab]}+{H}^{\mu\nu}_{e[ij,ab]}+{H}^{\mu\nu}_{f[ij,ab]}\ .
\end{align}
The physical amplitudes for the hyperon decay processes of our interest, specified in eq.~\eqref{eq.physical.processes}, can be expressed as linear combinations of the above SU(3)-Cartesian amplitudes~\eqref{eq.SU3.amplitude} as follows:
\begin{align}
 H^{\mu\nu}_{\Sigma^-\to\Sigma^+\ell^-\ell^-} &=  2\,H_{\frac{1-{\rm i}2}{\sqrt{2}}\frac{1-{\rm i}2}{\sqrt{2}},\frac{1+{\rm i}2}{\sqrt{2}}\frac{1+{\rm i}2}{\sqrt{2}}}^{\mu\nu}= \frac{1}{2}(H^{\mu\nu}_{11,11}-{\rm i}H^{\mu\nu}_{12,11}+\cdots)\ , \notag \\
 H^{\mu\nu}_{\Sigma^-\to p\,\,\ell^-\ell^-} &=  2\left( H_{\frac{1-{\rm i}2}{\sqrt{2}}\frac{4-{\rm i}5}{\sqrt{2}},\frac{1+i2}{\sqrt{2}}\frac{4+{\rm i}5}{\sqrt{2}}}^{\mu\nu}+ H_{\frac{1-{\rm i}2}{\sqrt{2}}\frac{4-{\rm i}5}{\sqrt{2}},\frac{4+{\rm i}5}{\sqrt{2}}\frac{1+{\rm i}2}{\sqrt{2}}}^{\mu\nu}\right)\ ,\notag\\
 H^{\mu\nu}_{\Xi^-\to \Sigma^+\ell^-\ell^-} &=  2 \left(H_{\frac{4-{\rm i}5}{\sqrt{2}}\frac{1-{\rm i}2}{\sqrt{2}},\frac{1+{\rm i}2}{\sqrt{2}}\frac{4+{\rm i}5}{\sqrt{2}}}^{\mu\nu}+ H_{\frac{4-{\rm i}5}{\sqrt{2}}\frac{1-{\rm i}2}{\sqrt{2}},\frac{4+{\rm i}5}{\sqrt{2}}\frac{1+{\rm i}2}{\sqrt{2}}}^{\mu\nu}\right)\ ,\notag\\
\label{eq:Amp_phys_Xpll}
 H^{\mu\nu}_{\Xi^-\to p\,\,\ell^-\ell^-} &=   2\,H_{\frac{4-{\rm i}5}{\sqrt{2}}\frac{4-{\rm i}5}{\sqrt{2}},\frac{4+{\rm i}5}{\sqrt{2}}\frac{4+{\rm i}5}{\sqrt{2}}}^{\mu\nu}\ .
\end{align}
In fact, the amplitude in eq.~\eqref{eq.SU3.amplitude}
is only responsible for the $u$-channel contribution. The $t$-channel contribution can be obtained with the help of crossing symmetry, as discussed section~\ref{sec:OneLoop_Amp} and appendix~\ref{app:cross}.

\section{Relations between the $t$- and $u$-channel structure functions}
\label{app:cross}
As illustrated in section~\ref{sec:OneLoop_Amp}, the $t$-channel hadronic tensor is obtainable from the $u$-channel one with the help of eq.~\eqref{eq.tu.hadronic.tensor}, subject to crossing symmetry. The $t$-channel hadronic tensor can be further decomposed as
\begin{align}
     {H}^t_{\mu\nu} =\sum_{i=1}^{34}\big[ V_i^t \mathcal{O}_{V,\mu\nu}^i+A_i^t \mathcal{O}_{A,\mu\nu}^i\big]\ ,
\end{align}
where the vector and axial-vector structure functions, $V_i^t$ and $A_i^t$ ($i=1,\cdots,34$) are related to the $u$-channel analogues $V_i^u$ and $A_i^u$. The relations between the vector structure functions $V_i^t$ and $V_j^u$ are given by 
\begin{align}
V^t_{1}&=V^u_{1}-\Delta_{B}(V^u_{2}+2V^u_{3})+2V^u_4\ ,  \quad
V^t_{2}=-V^u_{2}-2V^u_{3} \ , \quad V^t_3 = V^u_3 \ ,  
\notag \\
V^t_{4}&=\Delta_{B}V^u_{3}-V^u_{4}\ , \quad 
V^t_{5}=V^u_{5}-\Delta_{B}V^u_{14}+2(V^u_{31}+V^u_{32})\ , \notag\\
V^t_{6}&=V^u_{6}-\Delta_{B}V^u_{15}+2(V^u_{33}+V^u_{34})\ ,\quad
V^t_{7}=V^u_{8}-\Delta_{B}V^u_{17}+2(V^u_{32}+V^u_{33})\ , \notag\\
V^t_{8}&=V^u_{7}-\Delta_{B}V^u_{16}+2(V^u_{31}+V^u_{34})\ , \quad
V^t_{9}=V^u_{9}-\Delta_{B}V^u_{18}-2(V^u_{29}+V^u_{30}) \ , \notag\\
V^t_{10}&=V^u_{13}-\Delta_{B}V^u_{22}+2(V^u_{30}-V^u_{33})\ , \quad
V^t_{11}=V^u_{12}-\Delta_{B}V^u_{21}+2(V^u_{30}-V^u_{31})\ , \notag\\
V^t_{12}&=V^u_{11}-\Delta_{B}V^u_{20}+2(V^u_{29}-V^u_{32})\ , \quad
V^t_{13}=V^u_{10}-\Delta_{B}V^u_{19}+2(V^u_{29}-V^u_{34}) \ , \notag\\
V^t_{14}&= -V^u_{14}\ , \quad  V^t_{15} = -V^u_{15} \ , \quad V^t_{16} =-V^u_{17} \ , \quad V^t_{17} = -V^u_{16}\ ,\quad V^t_{18} = -V^u_{18} \ ,
\notag\\
V^t_{19} &= -V^u_{22}\ , \quad
V^t_{20} = -V^u_{21}\  , \quad  V^t_{21} = -V^u_{20}\ , \quad V^t_{22} = -V^u_{19}\ , \notag\\
V^t_{23}&=-2V^u_{3}+V^u_{24}+\Sigma_{B}V^u_{30}\ ,      \quad
V^t_{24}=2V^u_{3}+V^u_{23}+\Sigma_{B}V^u_{29}\ ,            \notag\\
V^t_{25}&=2V^u_{3}+V^u_{28}+\Sigma_{B}V^u_{32}\ ,            \quad
V^t_{26}=2V^u_{3}+V^u_{27}+\Sigma_{B}V^u_{34}\ ,            \notag\\
V^t_{27}&=-2V^u_{3}+V^u_{26}+\Sigma_{B}V^u_{33}\ ,            \quad
V^t_{28}=-2V^u_{3}+V^u_{25}+\Sigma_{B}V^u_{31}\ ,            \notag\\
V^t_{29} &= -V^u_{30}\ ,\quad  V^t_{30} = -V^u_{29}\ , \quad
V^t_{31} = -V^u_{32}\ ,  \notag \\
V^t_{32} &= -V^u_{31}\ , \quad V^t_{33} = -V^u_{34}\ , \quad V^t_{34} = -V^u_{33}\ ,\label{eq.tu.Vrelation}
\end{align}
with \(\Delta_B =  m_{2}-m_{1}\) and \(\Sigma_B = m_{2} + m_{1}\). Notice that the interchange of Mandelstam variables $u\leftrightarrow t$, corresponding to $k_1\leftrightarrow k_2$, is implied for the $V_i^u$'s on the right-hand side of eq.~\eqref{eq.tu.Vrelation}. The relations between the axial-vector structure functions $A_i^t$ and $A_j^u$ can be easily obtained by replacing $V\to A$ and $\Sigma_B\leftrightarrow \Delta_B$ in Eq.~\eqref{eq.tu.Vrelation}. For example, $A^t_{1}=A^u_{1}-\Sigma_{B}(A^u_{2}+2A^u_{3})+2A^u_{4}$. 

The crossing relations~\eqref{eq.tu.Vrelation} enable us to derive the $t$-channel hadronic structure functions from the $u$-channel ones, which are directly calculated from the one-loop diagrams in figure~\ref{fig.Feynman.Diagram}. Conversely, we have also directly computed the $t$-channel one-loop diagrams and checked the correctness of the obtained $t$/$u$-channel amplitudes with these relations.

\section{Explicit expressions of neutrinoless transition form factors}\label{app:TFF_expressions}

In $d$-dimensional spacetime, the one-loop $N$-point scalar integrals are defined by~\cite{Denner:2005nn} 
\begin{align}\label{eq.loopint}
T_{N} = &\kappa  \int \frac{{\rm d}^dk}{[k^2-\texttt{m}_{1}^2+{\rm i}\epsilon][(k+p_1)^2-\texttt{m}_{2}^2+{\rm i}\epsilon] \cdots[(k+p_{N-1})^2-\texttt{m}_{N}^2+{\rm i}\epsilon]}\ ,
\end{align}
where the prefactor $\kappa=(2\pi\mu)^{4-d}/({\rm i}\pi^2)$, and $\epsilon$ is an infinitesimal positive number. The renormalization scale $\mu$ has mass dimension, which is introduced to maintain the correct dimension of the integral. As usual, we employ the traditional notation for the one-loop integrals with $N\leq 4$:
\begin{align}
    T_1&= A_0(\texttt{m}_1)\ ,\\
    T_2&=B_0(p_1^2,\texttt{m}_1^2,\texttt{m}_2^2)\ ,\\
    T_3&= C_0(p_1^2,(p_2-p_1)^2,p_2^2,\texttt{m}_1^2,\texttt{m}_2^2,\texttt{m}_3^2)\ ,\\
    T_4&= D_0(p_1^2,(p_2-p_1)^2,(p_3-p_2)^2,p_3^2,p_2^2,(p_3-p_1)^2,\texttt{m}_1^2,\texttt{m}_2^2,\texttt{m}_3^2,\texttt{m}_4^2)\ .
\end{align}
Furthermore, to make our final results of the one-loop neutrinoless TFFs (i.e., \eqref{eq:S_SSll}, \eqref{eq:S_Spll} and \eqref{eq:S_Xpll}) compact, the following abbreviations for the specific integrals are defined: 
\begin{align}
A_0&=A_0(m^2)\ ,  \qquad B_0^1=B_0(m^2,m^2,M_{K}^2)\ , \qquad B_0^2=B_0(m^2,m^2,M_{\pi}^2)\ ,  \notag\\
    B_0^3&=B_0(m_{\ell}^2,0,M_{K}^2)\ , \quad B_0^4=B_0(m_{\ell}^2,0,M_{\pi}^2)\ , \quad B_0^5=B_0(\frac{\Sigma_m-s}{2},0,m^2)\ ,  \notag\\
    C_0^1&=C_0(m_{\ell}^2,s,m_{\ell}^2,0,M_K^2,M_K^2)\ , \qquad C_0^2=C_0(m_{\ell}^2,s,m_{\ell}^2,0,M_K^2,M_{\pi}^2)\ ,   \notag\\
    C_0^3&=C_0(m_{\ell}^2,s,m_{\ell}^2,0,M_{\pi}^2,M_{\pi}^2)\ , \qquad C_0^4=C_0(\frac{\Sigma_m-s}{2},m^2,m_{\ell}^2,0,m^2,M_K^2)\ , \notag\\
    C_0^5&=C_0(\frac{\Sigma_m-s}{2},m^,m_{\ell}^2,0,m^2,M_{\pi}^2)\ ,  \notag\\
    D_0^1&=D_0(\frac{\Sigma_m-s}{2},m^2,s,m_{\ell}^2,m_{\ell}^2,m^2,0,m^2,M_K^2,M_K^2)\ , \notag\\
    D_0^2&=D_0(\frac{\Sigma_m-s}{2},m^2,s,m_{\ell}^2,m_{\ell}^2,m^2,0,m^2,M_K^2,M_{\pi}^2)\ , \notag\\
    D_0^3&=D_0(\frac{\Sigma_m-s}{2},m^2,s,m_{\ell}^2,m_{\ell}^2,m^2,0,m^2,M_{\pi}^2,M_{\pi}^2)\ , \label{eq.ABCD}
\end{align}
with $\Sigma_m=2m^2+2m_{\ell}^2$. 

In what follows, explicit expressions of the one-loop TFFs, defined in Eq~\eqref{eq:TFFS}, for the four typical physical processes $\Sigma^- \to p\ell^-\ell^-$, $\Sigma^- \to \Sigma^+ \ell^-\ell^-$, $\Xi^- \to p\ell^- \ell^-$ and $\Xi^- \to \Sigma^+ \ell^-\ell^-$ are listed. Note that we use the SU(3) chiral limit mass $m$, instead of their physical masses, for the octet baryons in our one-loop results. Specifically, for the $\Delta S=0$ process $\Sigma^- \to \Sigma^+\ell^-\ell^-$, the one-loop TFF reads
\begin{align}\label{eq:S_SSll}
\hspace{-0.3cm}
\mathcal{S}^{\rm loop}_{\Sigma^-\to \Sigma^+\ell\ell}(s)&=-\frac{m(D^2-3F^2)}{3(16 m^2 m_\ell^2-s^2)(4m^2-s)} \Bigg[C_0^3s(-2M_{\pi}^2+s)(16m^2m_{\ell}^2-s^2)  \notag\\
&+ 2D_0^3 m^2(-2 M_{\pi}^2+s)^2(16m^2m_{\ell}^2-s^2) + 8B_0^2 m^2s(4m^2-s)  \notag\\
&- 2B_0^4 s^2(4m^2-s) + 4C_0^5\bigg(M_{\pi}^2 s^3 +8 m^4m_\ell^2(16 M_{\pi}^2-3s)-8 m^4M_{\pi}^2s \notag\\
&+m^2s (-4sM_{\pi}^2+s^2+-24m_\ell^2M_{\pi}^2+2sm_\ell^2) \bigg) \Bigg] +\frac{A_0^1(D^2-3-3F^2)m(d-2)}{3(\Sigma_m-s)}   \notag\\
& -\frac{B_0^5m}{6(16 m^2 m_\ell^2-s^2)(\Sigma_m-s)} \Bigg[-3(1+D^2-3F^2)(2 m_\ell^2-s)s^2(2+d)  \notag\\
& +32 m^4\Big((4m_\ell^2(2+d)-s)(D^2-3F^2)+12m_\ell^2\Big)   \notag\\
&+8 m^2\Big(-3s^2-s^2(D^2-3F^2)(d-1) -6m_\ell^2s(1+D^2-3F^2)(4+d) \notag\\
&+4m_\ell^4(1+D^2-3F^2)(10+3d)+12m_\ell^2s-16m_\ell^4\Big) \Bigg]\ .
\end{align}
For the $\Delta S=1$ processes $\Sigma^- \to p\ell^-\ell^-$ and $\Xi^- \to \Sigma^+\ell^-\ell^-$, one has
\begin{align}\label{eq:S_Spll}
\hspace{-0.3cm}
\mathcal{S}^{\rm loop}_{\Sigma^-\to p\ell\ell}(s)&=\mathcal{S}^{\rm loop}_{\Xi^-\to \Sigma^+\ell\ell}(s)\notag\\
&=-\frac{m(D^2-3F^2)}{3(16m^2m_\ell^2-s^2)(4m^2-s)} 
\Bigg[2D_0^2 m^2 (M_{K}^2+M_{\pi}^2-s)^2 (16 m^2 m_\ell^2-s^2) \notag\\
&-C_0^2(M_{K}^2+M_{\pi}^2-s)s (16 m^2 m_\ell^2-s^2) + 2C_0^4 \Big(8 m^4 \big(m_\ell^2(2 M_{\pi}^2-3s)     \notag\\
& +M_{K}^2(14 m_\ell^2-s)\big) +M_{K}^2 s^3 + m^2 s\big(-3 M_{K}^2(8 m_\ell^2+s)+s(2m_\ell^2-M_{\pi}^2+s)\big)\Big) \notag\\
& + 2C_0^5 \Big(M_{\pi}^2 s^3 +8 m^4 \big(2 M_{K}^2 m_\ell^2+m_\ell^2(14 M_{\pi}^2-3s)-M_{\pi}^2 s\big) +m^2 s\big(s^2-sM_{K}^2        \notag\\
& -3sM_{\pi}^2 +m_\ell^2(-24 M_{\pi}^2+2s)\big)\Big) \Bigg]  +\frac{A_0^1 (D^2-3-3F^2)m(d-2)}{3(\Sigma_m-s)}\notag\\
&+\frac{B_0^5m}{6(16m^2m_\ell^2-s^2)(\Sigma_m-s)} \Big[3(1+D^2-3F^2)(2 m_\ell^2-s)s^2(2+d) \notag\\
&-32 m^4\Big(-(D^2-3F^2)s+4 m_\ell^2\big(3+(D^2-3F^2)(2+d)\big)\Big) \notag\\
&-8 m^2\Big(s^2\big(-3+(3F^2-D^2)(d-1)\big)-6 m_\ell^2 s(1+D^2-3F^2)(4+d) \notag\\
&-4 m_\ell^4(1+D^2-3F^2)(10+3d)+12 m_\ell^2 s-16 m_\ell^4 \Big)     \Big] \notag\\
&-\frac{m(D^2-3F^2)s}{3(16m^2m_{\ell}^2-s^2)}\big(4(B_0^1+B_0^2)m^2 + (B_0^3+ B_0^4)s \big)  \ .
\end{align}
Finally, for the $\Delta S=2$ decay of $\Xi^- \to p \ell\ell$, we obtain
\begin{align}\label{eq:S_Xpll}
\hspace{-0.3cm}
\mathcal{S}^{\rm loop}_{\Xi^-\to p\ell\ell}(s)&=-\frac{m(D^2-3F^2)}{3(16 m^2 m_\ell^2-s^2)(4m^2-s)}\Bigg[-2D_0^1 m^2 (-2 M_K^2+s)^2(-16 m^2 m_\ell^2+s^2)  \notag\\
&+ 4C_0^4\Big(M_K^2 s^3 + 8 m^4 \big(M_K^2(16 m_\ell^2-s)-3 m_\ell^2 s\big) 
+ m^2 s\big(s(2 m_\ell^2+s)     \notag\\
&-4 M_K^2(6 m_\ell^2+s)\big)  \Big) +C_0^1 s(-2 M_K^2+s)(16 m^2 m_\ell^2-s^2)  \Bigg]      \notag\\
&+\frac{A_0^1 (D^2-3-3F^2)m(d-2)}{3(\Sigma_m-s)}-\frac{m(D^2-3F^2)(8B_0^1m^2s-2B_0^3s^2)}{3(16m^2m_{\ell}^2-s^2)} \notag\\
&+ \frac{mB_0^5}{3(16 m^2 m_\ell^2-s^2)(\Sigma_m-s)} \Big[ -3(1+D^2-3F^2)(2 m_\ell^2-s)s^2(2+d) \notag\\
& +32 m^4\Big(-(D^2-3F^2)s
+4 m_\ell^2\big(3+D^2(2+d)-3F^2(2+d)\big)\Big) \notag\\
&+8 m^2\Big(-3s^2-s^2(D^2-3F^2)(d-1) -6m_\ell^2 s(1+D^2-3F^2)(4+d) \notag\\
&+12m_\ell^2 s +4m_\ell^4(1+D^2-3F^2)(10+3d)-16m_\ell^4\Big)
\Big]\ .
\end{align}
In our numerical computation, the one-loop scalar integrals are evaluated using both \texttt{LoopTools}~\cite{Hahn:1998yk} and \texttt{AMFlow}~\cite{Liu:2022chg}, while the relevant UV divergences are removed by applying the $\overline{\rm MS}$-1 subtraction scheme. For the above one-loop TFFs, the PCB term that needs to be subtracted reads
\begin{align}
\mathcal{S}^{\rm PCB} &= -\frac{(d-2)[3d-3+(7+d)(D^2-3F^2)]A_0(m^2)}{6(d-3)m}\ .
\end{align}

\bibliographystyle{JHEP}
\bibliography{biblio}
\end{document}